\documentclass[journal,onecolumn,12pt]{IEEEtran}
\usepackage{diagbox}
\usepackage{graphicx}
\usepackage{epstopdf}
\usepackage{threeparttable}
\usepackage{amsmath,amssymb}
\usepackage{arydshln}
\usepackage{dsfont}
\usepackage{color}
\usepackage{cases}
\usepackage{bm}
\usepackage{amsfonts}
\usepackage[font=footnotesize]{caption}
\usepackage{indentfirst}
\usepackage{cite}
\usepackage{caption}
\usepackage{algorithm}
\usepackage{algpseudocode}
\usepackage{booktabs}
\usepackage{subfigure}
\usepackage{multirow}
\usepackage{amsthm}
\usepackage{setspace}
\newcommand{\sys}{}
\newcommand{\sd}{\mathbf}

\usepackage{hyperref}

\theoremstyle{remark}
\newtheorem{remark}{Remark}
\theoremstyle{}
\newtheorem{theorem}{Theorem}
\theoremstyle{}

\newtheorem{lemma}{Lemma}
\theoremstyle{}
\newtheorem{definition}{Definition}
\theoremstyle{remark}
\newtheorem{example}{Example}

\theoremstyle{definition}

\makeatletter
\newcommand{\tabcaption}{\def\@captype{table}\caption}

\makeatletter
 \allowdisplaybreaks[4]

\ifCLASSINFOpdf
\else
\fi

\definecolor{newcolor}{rgb}{0.5,0,1}

\begin{document}

\title{Robust, Private and Secure Cache-aided Scalar Linear Function Retrieval from Coded Servers}

\author{
Qifa Yan~\IEEEmembership{Member,~IEEE,} and
Daniela Tuninetti~\IEEEmembership{Fellow,~IEEE.}
\thanks{A short version of this paper will appear in the 2021 IEEE Symposium on Information Theory (ISIT).}
\thanks{Q. Yan is with the Information Security and National Computing Grid Laboratory, Southwest Jiaotong University, Chengdu 611756, China (email: qifayan@swjtu.edu.cn). The work was done whole when Dr. Yan was  with the Electrical and Computer Engineering Department of the University of Illinois Chicago, Chicago, IL 60607, USA. 
}
\thanks{ D. Tuninetti is with the Electrical and Computer Engineering Department
of the University of Illinois Chicago, Chicago, IL 60607, USA (e-mail:
danielat@uic.edu).}
\thanks{
 This work was supported in part by NSF Award 1910309.
}
}
\maketitle
\pagestyle{empty}  
\thispagestyle{empty} 
\IEEEpeerreviewmaketitle

\begin{abstract}
This work investigates a system where each user aims to retrieve a scalar linear function of the files of a library, which are Maximum Distance Separable coded and stored at multiple distributed servers.
The system needs to guarantee \emph{robust decoding} in the sense that each user must decode its demanded function with signals received from any subset of servers whose cardinality exceeds a threshold. In addition,
(a) the content of the library must be kept secure from a wiretapper who obtains all the signals from the servers;
(b) any subset of users together can not obtain any information about the demands of the remaining users; and
(c) the users' demands must be kept private against all the servers even if they collude.
Achievable schemes are derived by modifying existing Placement Delivery Array (PDA) constructions, originally proposed for single-server single-file retrieval coded caching systems without any privacy or security or robustness constraints.
It is shown that the PDAs describing the original Maddah-Ali and Niesen's coded caching scheme result in a load-memory tradeoff that is optimal to within a constant multiplicative gap, except for the small memory regime when the number of file is smaller than the number of users. As by-products, improved order optimality results are derived for three less restrictive systems in all parameter regimes.     
\end{abstract}


\section{Introduction}
\label{sec:intro}

Coded caching, introduced by Maddah-Ali and Niesen (MAN)~\cite{Maddah-Ali2014fundamental}, is a technique to reduce the peak-time communication load across a bottleneck shared link by leveraging the multicast opportunities created by content pre-stored at users' local caches. The model consists of a single server, multiple users, and two phases. In the \emph{placement phase},  the users' caches are populated without the knowledge of their future demands. In the \emph{delivery phase}, when users' demands are revealed, the server satisfies them by transmitting coded packets over the shared link.
For a system with $N$ files and $K$ users, the MAN scheme achieves the optimal load-memory tradeoff among all uncoded placement schemes when $N\geq K$~\cite{Kai2020Index}, and for $N <K$ after removing some redundant transmissions~\cite{Yu2019ExactTradeoff}. Recently, it was showed that allowing the users to demand arbitrary linear combinations of the files does not increase the load compared to the case single file retrieval, at least under uncoded placement~\cite{Kai2020LinearFunction}.

Content security,  demand (both user- and server-side) privacy, and robustness are critical aspects of practical systems.

\paragraph*{Content Security} 
In~\cite{Security}, the content of the library must be protected against an external wiretapper who obtains the signals transmitted during the delivery phase. The key idea in~\cite{Security} is that users cache the same content as in the MAN scheme~\cite{Maddah-Ali2014fundamental}, and in addition also share some \emph{security keys} for the part of the files that were not cached in the MAN scheme. The latter is done in a structured way so that each user
can retrieval all the multicast signals it needs to decode.

\paragraph*{User-side Demand Privacy} 
Schemes that guarantee user privacy, that is, no user can infer the demand of another user after the delivery phase, were proposed in~\cite{Kai2019Private}. In particular, user privacy can be guaranteed by adding virtual users~\cite{Kai2019Private,Kamath2019}.
We investigated user privacy against colluding users in~\cite{Y:D:Privacy}, for both single file retrieval and scalar linear function retrieval, where we imposed that any subset of users must not obtain any information about the demands of other users even if they exchange the content in their caches.
The key idea in~\cite{Y:D:Privacy} is that, in addition to the cached contents as in the MAN scheme~\cite{Maddah-Ali2014fundamental}, each user also privately caches some  \emph{privacy keys}, which are composed as random linear combinations of the parts of the files that were not cached in the MAN scheme. The demands are added by the same coefficients used to generate the privacy keys, so that each user can decode its demanded files with the privacy keys. 

\paragraph*{Content Security \& User-side Demand Privacy}
We investigated simultaneous content Security and user demand Privacy for scalar Linear Function Retrieval (SP-LFR) in~\cite{Y:D:SP-LFR}, where we designed a \emph{key superposition} scheme to guarantee both conditions at once by superposing (i.e., sum together) the security keys and privacy keys. We showed that the load-memory tradeoff in this case is the same as in the setup with only content security guarantees. The idea of key superposition was incorporated into the framework of Placement Delivery Array (PDA), which was known to depict both placement and delivery phases in a single array for coded caching systems with neither security or privacy constraint~\cite{Yan2017PDA}.  The advantage of the PDA framework is that low subpacketization schemes can be obtained directly from existing PDA constructions, such as the ones in~\cite{Yan2017PDA,PDA:bipartite,PDA:a,PDA:b,PDA:c}.

\paragraph*{Server-side Demand Privacy} 
Server-side demand privacy has been thoroughly investigated for the case of multiple servers and a single user, which is known as the Private Information Retrieval (PIR) problem~\cite{ChorPIR95}. The capacity of PIR has been characterized in~\cite{HusPIRCapacity} for single file retrieval, in~\cite{HuaPIRComputation} for scalar linear function retrieval, and in~\cite{ColludingServerPIR} or single file retrieval and colluding servers.  PIR with a cache-aided user was investigated in~\cite{Tandon17,UlukusPIR2018,UlukusPIR2019,UlukusPIR2019Dec}. Recently, the PIR setting has been extended so as to include multiple cache-aided users in~\cite{XiangZhang, XiangZhang02}, where techniques from coded caching and PIR were combined to derive achievable scheme that are provably optimal to within a constant gap.

\paragraph*{MDS Coded Servers and Decoding Robustness}
Since node failures and erasures commonly arise in storage systems, redundancy is desirable \cite{Dimarkis2011}.
Maximum Distance Separable (MDS) codes are often used to code the data stored across servers.
The advantage of MDS coded servers is that it saves storage while allowing unresponsive servers. 
PIR from MDS-coded servers has been investigated in~\cite{PIR_Coded_Servers,Jinbao2020,ZhouMDS_PIR}, and the capacity was charactered in \cite{PIR_Coded_Servers}. The schemes in~\cite{Jinbao2020,ZhouMDS_PIR} have almost optimal sub-packetization among all schemes achieving the smallest download rate. The PIR schemes in~\cite{RobustPIR} have asymptotically optimal download rate when any number of unresponsive servers not exceeding some threshold show up.

\subsection{Contributions and Paper Organization}

\begin{figure}
  \centering
  \includegraphics[width=0.45\textwidth]{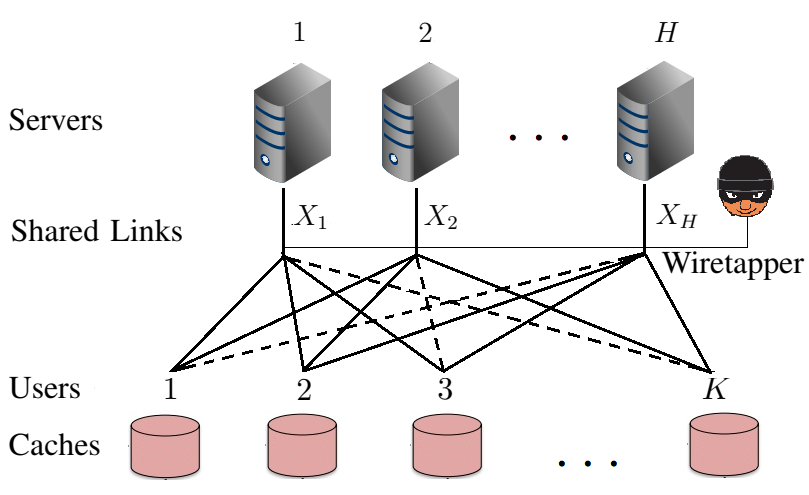}
  \caption{System model}\label{fig:system}
\vspace{-10pt}
\end{figure}

In this paper, we combine all the above mentioned requirements in a system whose model is depicted in Fig.~\ref{fig:system}.
The model consists of $H$ servers,  $N$ files, and  $K$ users.
Each of the $N$ files is stored, as an $(H,L)$ MDS coded version\footnote{%
An $(H,L)$ MDS code encodes $L$ information packets into $H$ coded packets, with the property that upon obtaining any $L$ (out of $H$) coded packets one can recover the $L$ information packets.
}, at all servers. 
Each server is connected to all the users via a dedicated  shared link, but may not be able to reach all the users. 
The novel aspect of this work is to \emph{design coded caching schemes that are robust to some servers' unavailability}, that is, each user must be able to retrieve an arbitrary scalar linear function of the files from the signals obtained from an arbitrary subset of $L$ servers (out of $H$ servers).   The security~\cite{Security}, user-side privacy~\cite{Y:D:Privacy} and server-side privacy~\cite{ColludingServerPIR} conditions are also imposed. We refer to this model as a Robust Secure and (server- and user-side) Private Linear Function Retrieval (RSP-LFR) problem.

Our key idea on how to guarantee all those conditions simultaneously is to extend the key superposition scheme in~\cite{Y:D:SP-LFR}. In particular, the technique of superposing user-side privacy and security keys is used in the placement phase, while in the delivery phase, the multicast signals are created in the MDS code domain, where the MDS coded version of the keys are added to the MDS coded multicast signals. 
Robustness is guaranteed by the linearity property of the MDS code. 
Security and (server- and user-side) privacy are guaranteed since each transmitted signal is accompanied by an appropriate MDS coded key.

Our main contributions for the proposed RSP-LFR model are as follows.
\begin{enumerate}

\item We propose a procedure to obtain a RSP-LFR scheme from a given PDA, so that low-subpacketization RSP-LFR schemes can be easily obtained from various existing PDA constructions~\cite{Yan2017PDA,PDA:bipartite,PDA:a,PDA:b,PDA:c}. Interestingly, with the same PDA, compared to the single server SP-LFR system in \cite{Y:D:SP-LFR},  the  achieved memory size is the same, but  the load is scaled by a factor $H/L$, i.e., the inverse of the rate of the MDS code used to encode the library files. 

\item Following the proposed procedure, RSP-LFR schemes based on the PDAs that describe the original MAN scheme in~\cite{Maddah-Ali2014fundamental} (MAN-PDAs) are proved to achieve the best load-memory tradeoffs among all PDA-based RSP-LFR schemes. Moreover, we show that they have the smallest subpacketization among all PDA based schemes achieving the same load-memory pairs.

\item The load-memory tradeoff achieved by MAN-PDAs is proved to be to within a constant multiplicative gap from the optimal load-memory tradeoff, except for the regime of small memory and less files than users.

\item For three less restrictive models, where some conditions are dropped, we propose schemes for the corresponding setups that improve the load-memory tradeoffs of the novel MAN-PDA-based RSP-LFR scheme. The idea for improving the tradeoff in less restrictive models is as follows. In the case where security is not imposed, security keys can be removed, and hence, some signals in the delivery phase became redundant and can be removed akin to~\cite{Yu2019ExactTradeoff,Kai2020LinearFunction,Y:D:SP-LFR}. Moreover, those improved schemes are shown to be optimal to within a constant multiplicative gap in their respective setups in all parameter regimes, and the gap is lower than previously known schemes.

\end{enumerate}

The rest of this paper is organized as follows.
Section~\ref{sec:model} gives the formal problem definition.
Section~\ref{sec:PDA:example} reviews the PDA framework and gives an illustrative example.
Section~\ref{sec:main} summarizes our main results, where the proof details are deferred to Sections \ref{sec:DSP-LFR:scheme}--\ref{sec:degraded}.
Section~\ref{sec:numerical} presents some numerical results.
Section~\ref{sec:conclusion} concludes the paper.

\subsection{Notation Convention} 
In this paper,  
$\mathbb{N}^+$ denotes the set of positive integers;
$\mathbb{F}_q$ and $\mathbb{F}_q^n$ denote the finite field of cardinality  $q$, for some prime power $q$, and the $n$-dimensional vector space over $\mathbb{F}_q$, respectively. 
For two integers $m,n$ such that $m\leq n$, we use $[m:n]$ to denote the set of the first positive integers $\{m,\ldots,n\}$; $[1:n]$ is also denoted by $[n]$ for short. 
We use $X_{\mathcal{A}}$ to denote the tuple composed of $\{X_i:i\in\mathcal{A}\}$ for some integer set $\mathcal{A}$, where the elements are ordered increasingly, e.g., $X_{[3]}=(X_1,X_2,X_3)$. 
For variables with two or more indices, e.g., $X_{i,j}$, we use $X_{\mathcal{A},\mathcal{B}}$ to denote the tuple $\{X_{i,j}: i\in\mathcal{A},j\in\mathcal{B} \}$, where the elements are listed in lexicographical order, e.g. $X_{[3],[2]}=(X_{1,1},X_{1,2},X_{2,1},X_{2,2},X_{3,1},X_{3,2})$.

\section{System Model}
\label{sec:model}

Let $N,K,L,H$ be positive integers satisfying $L\leq H$.
The $(N,K,L,H)$ RSP-LFR system, illustrated in Fig.~\ref{fig:system}, consists of $H$ servers (denoted by $1,\ldots,H$), where each server is connected to $K$ users (denoted by $1,\ldots,K$) via a dedicated shared-link.  
A file library of $N$ files (denoted by $W_1,\ldots,W_{N}\in\mathbb{F}_q^B$) are stored at the $H$ servers in the form of an $(H,L)$ MDS code as follows, where $B$ denotes the file length.  
Each file $W_n, n\in[N],$ is composed of $L$ equal-size subfiles $W_{n,1},\ldots,W_{n,L}\in\mathbb{F}_q^{B/L}$ and is encoded into $H$ coded subfiles $\overline{W}_{n,1},\ldots,\overline{W}_{n,H}\in\mathbb{F}_q^{B/L}$ with a given $(H,L)$ MDS code with generator matrix
\begin{IEEEeqnarray}{c}
G=\left[\begin{array}{ccc}
g_{1,1}&\ldots&g_{1,H}\\
\vdots&\ddots &\vdots\\
g_{L,1}&\ldots&g_{L,H}
\end{array}
\right],
\end{IEEEeqnarray}
that is, the coded subfiles are given by 
\begin{IEEEeqnarray}{rCl}\label{eqn:MDS:code}
\lefteqn{(\overline{W}_{n,1},\ldots,\overline{W}_{n,H})}\notag\\
&=&\Big(\sum_{l\in[L]}g_{l,1}W_{n,l},\ldots,\sum_{l\in[L]}g_{l,H}W_{n,l}\Big),~\forall n\in[N].\IEEEeqnarraynumspace
\end{IEEEeqnarray}
The $N$ files are mutually independent and uniformly distributed over $\mathbb{F}_q^B$, that is, 
\begin{subequations}\label{eqn:file:entropy}
\begin{IEEEeqnarray}{rCl}
H(W_1)&=&\ldots =H(W_N)=B,\\
H(W_1,\ldots,W_{N})&=&H(W_1)+\ldots+H(W_{N}).\IEEEeqnarraynumspace
\end{IEEEeqnarray}
\end{subequations}
Therefore, each subfile or coded subfile is uniformly distributed over $\mathbb{F}_q^{B/L}$. 
Server $h\in[H]$ stores the $h$-th coded subfile of each file, i.e., 
\begin{IEEEeqnarray}{c}
\overline{W}_{[N],h} := (\overline{W}_{1,h}, \ldots, \overline{W}_{N,h}),~\forall h\in[H].
\end{IEEEeqnarray}

For notational simplicity, for a vector $\sd{a}=(a_1,\ldots,a_N)^\top \in \mathbb{F}_q^{N}$, we denote the scalar (i.e., operations are meant element-wise across files) linear combination of the files or (coded) subfiles for all $l\in[L]$ and $h\in[H]$  as
\begin{subequations}
\begin{IEEEeqnarray}{rCl}
W_{\sd{a}}  &:=&\sum_{n\in[N]}a_n W_n, \\
W_{\sd{a},l}& := &\sum_{n\in[N]}a_n W_{n,l},
\\
\overline{W}_{\sd{a},h} &:=& \sum_{n\in[N]}a_n\overline{W}_{n,h} = \sum_{l\in[L]}g_{l,h}W_{\sd{a},l}.
\label{eq:MSDofLC}
\end{IEEEeqnarray}
\label{eqn:LinearFiles}
\end{subequations}
Notice that, $W_{\sd{a}},W_{\sd{a},l},\overline{W}_{\sd{a},h}$ are linear in $\sd{a}$, e.g., for any $u,v\in\mathbb{F}_q$ and $\sd{a},\sd{b}\in\mathbb{F}_q^N$, $W_{u\sd{a}+v\sd{b}}=uW_{\sd{a}}+vW_{\sd{b}}$.
Moreover, since $\overline{W}_{n,[H]}:=(\overline{W}_{n,1},\ldots,\overline{W}_{n,H})$ is the MDS coded version of $W_{n,[L]}:=(W_{n,1},\ldots,W_{n,L}), \forall n\in[N],$ by linearity we have that $\overline{W}_{\sd{a},[H]}:=(\overline{W}_{\sd{a},1},\ldots,\overline{W}_{\sd{a},H})$ is the MDS coded version of $W_{\sd{a},[L]}:=(W_{\sd{a},1},\ldots,W_{\sd{a},L}), \forall\, \sd{a} \in \mathbb{F}_q^{N}$, as in~\eqref{eq:MSDofLC}.

The system operates in two phases as follows.

\paragraph*{Placement Phase} 
The servers can communicate with each other, and all users can access all servers. To ensure the security condition in~\eqref{eqn:security}, the servers share some randomness $V$ from some finite alphabet $\mathcal{V}$. 
Each user~$k\in[K]$ generates some random variable $P_k$ from some finite alphabet $\mathcal{P}_k$ and cache some content $Z_k$ as a function of $P_k$, $V$ and the file library $W_{[N]}$.
Let the cached content be
\begin{IEEEeqnarray}{c}
Z_k := \varphi_k(P_k,V,W_{[N]}) \in \mathbb{F}_q^{\lfloor MB\rfloor}, \forall\, k\in[K],
\end{IEEEeqnarray}
for some encoding functions 
$ 
\varphi_k:\mathcal{P}_k\times\mathcal{V}\times \mathbb{F}_q^{NB}\mapsto \mathbb{F}_q^{\lfloor MB\rfloor}, \ \forall\, k\in[K].
$ 
The quantity $M$ is referred to as \emph{memory size}. 
The encoding functions $\varphi_1,\ldots,\varphi_K$ are known by the servers, but the randomness $P_1,\ldots,P_K$ are kept private by the corresponding users.

\paragraph*{Delivery Phase} 
Each user~$k\in[K]$ generates a demand $\sd{d}_k=(d_{k,1},\ldots,d_{k,N})^\top\in\mathbb{F}_q^N$, meaning it is interested in retrieving the linear combination of the files $ W_{\sd{d}_k}$.
The following random variables 
are independent
\begin{subequations}
\begin{IEEEeqnarray}{rCl}
&&\lefteqn{H(\sd{d}_{[K]},W_{[N]},P_{[K]},V)
=}\\
&&\sum_{k\in[K]} \! H(\sd{d}_k)
+ \sum_{n\in[N]} \! H(W_n)
+ \sum_{k\in[K]} \! H(P_k)
+ H(V).\IEEEeqnarraynumspace
\label{eqn:ind:dpw}
\end{IEEEeqnarray}
\end{subequations}

User~$k\in[K]$ generates queries $Q_{k,[H]} := (Q_{k,1},\ldots,Q_{k,H})$ as
\begin{IEEEeqnarray}{c}
  Q_{k,h} := \kappa_{k,h} (\sd{d}_k,Z_k)\in\mathbb{F}_q^{\ell_{k,h}},\forall\, h\in[H],
\end{IEEEeqnarray}
for some query functions $\kappa_{k,h}:\mathbb{F}_q^N\times \mathbb{F}_q^{\lfloor MB\rfloor}\mapsto \mathbb{F}_q^{\ell_{k,h}}$, where $\ell_{k,h}$ is the length of the query $Q_{k,h}$.
If any randomness is needed in the queries, it has to be stored in the cache.
Then user~$k\in[K]$ sends the query $Q_{k,h}$ to server $h\in[H]$. 

Upon receiving the queries from all the users, server $h\in[H]$ creates a signal $X_h$ as
\begin{IEEEeqnarray}{c}
X_h := \phi_h(V,Q_{[K],h},\overline{W}_{[N],h}) \in \mathbb{F}_q^{\lfloor R_hB\rfloor}, \forall h\in[H],\IEEEeqnarraynumspace
\end{IEEEeqnarray}
for some encoding function 
$\phi_h:\mathcal{V}\times \mathbb{F}_q^{\sum_{k\in[K]}\ell_{k,h}}\times \mathbb{F}_q^{\frac{NB}{L}}\mapsto \mathbb{F}_q^{\lfloor R_hB\rfloor}$. 
The quantity $R_h, h\in[H],$ is referred to as the \emph{load of server $h$}. The \emph{(total) load} of the system is defined as
\begin{IEEEeqnarray}{c}
R := \sum_{h\in[H]}R_h.
\end{IEEEeqnarray}

An 
RSP-LFR scheme must satisfy the following conditions for all demands $\sd{d}_1,\ldots,\sd{d}_K\in\mathbb{F}_q^N$. 
\begin{subequations}
\begin{IEEEeqnarray}{l}
 \textrm{[Robust Correctness]}: \ H(W_{\sd{d}_k}\,|\,X_{\mathcal{L}},\sd{d}_k,Z_k)=0,
  \notag\\
  \qquad\qquad\qquad\qquad\qquad \forall\, k\in[K],\mathcal{L}\subseteq[H]:|\mathcal{L}|=L,
  \label{eqn:correctness}
  \\
  \textrm{[Security]}: \quad\quad I(W_{[N]};X_{[H]})=0,
  \label{eqn:security}
  \\
  \textrm{[User-side Privacy]}: ~I(\sd{d}_{[K]\backslash\mathcal{S}};Z_{\mathcal{S}},X_{[H]},\sd{d}_{\mathcal{S}},\,|\,W_{[N]})=0,\notag\\
~~~~~~~~~~~~~~~~~~~~~~~~~~~~~~~~~~~\forall\, \mathcal{S}\subseteq[K]:\mathcal{S}\neq \emptyset, 
  \label{eqn:user:privacy}
  \\
  \textrm{[Server-side Privacy]}: \notag\\\quad\quad\quad\quad\quad \quad I(\sd{d}_{[K]};Q_{[K],[H]},\overline{W}_{[N],[H]},V)=0,
  \label{eqn:server:privacy}
\end{IEEEeqnarray}
\label{eq:allconditions}
\end{subequations}
\paragraph*{Objective} A memory-load pair $(M,R)\in[1,N]\times \mathbb{R}^{+}$ is said to be $B$-achievable if, for any $\epsilon>0$,  there exists a scheme satisfying all the conditions in~\eqref{eq:allconditions} with memory size less than $M+\epsilon$, load less than $R+\epsilon$ with file-length $B$.  The main objective of this paper is to characterize the  optimal load-memory tradeoff of the system, defined as
\begin{IEEEeqnarray}{l}
R^*(M):=
  \inf_{B\in\mathbb{N}^+}
\big\{R\,:\,(M,R)~\textnormal{is $B$-achievable}\big\}.
\IEEEeqnarraynumspace
\label{RM:star}
\end{IEEEeqnarray}
Throughout this paper, we consider the case $N\geq 2$, since demand privacy is impossible for $N=1$ (i.e., there is only one possible file to be demanded).

For a given scheme, we are also interested in its subpacketization level, which is defined as the number of packets each file has to be partitioned into in order to implement the scheme.

\begin{remark}[Implications of the conditions in~\eqref{eq:allconditions}]
The constrains in~\eqref{eq:allconditions} imply the following.
\begin{enumerate}
   \item The robust correctness condition in~\eqref{eqn:correctness} guarantees that each user can correctly decode its required scalar linear function by receiving any $L$-subsets of the transmitted signals. Since each user decodes independently, the available subset of signals $\mathcal{L}$ need not to be same across the users. 
   \item The security condition in~\eqref{eqn:security} guarantees that a wiretapper, who is not a user in the system and observes all the delivery signals, can not obtain any information about the contents of the library files.
It was proved in~\cite[Appendix A]{Y:D:Privacy} that the conditions in~\eqref{eqn:security} and~\eqref{eqn:user:privacy} imply
\begin{IEEEeqnarray}{l}
I(W_{[N]},\sd{d}_{[K]};X_{[H]})=0,
\label{eqn:security:XW}
\end{IEEEeqnarray}
that is, the wiretapper having access to $X_{[H]}$ in fact can not obtain any information on both the files and the demands of the users.
   \item The user-side privacy condition in~\eqref{eqn:user:privacy} guarantees that any subset of users who exchange their cache contents cannot jointly learn any information on the demands of the other users, regardless of the file realizations.

   \item The server-side privacy condition in~\eqref{eqn:server:privacy} guarantees that the servers can not obtain any information on the demands of the users, even if all the servers collude by exchanging their stored contents. 

 \end{enumerate}
\end{remark}

\begin{remark}[Minimum memory size]\label{remark:M}
It was proved in~\cite{Security} that, in order to guarantee the correctness condition in~\eqref{eqn:correctness} and the security condition in~\eqref{eqn:security} simultaneously, the memory size $M$ has to be no less than one.
Thus the load-memory tradeoff is defined for $M\in[1,N]$.
\end{remark}



\begin{remark}[Comparison with~\cite{XiangZhang}] In the case $L=1$ and $G=[1,1,\ldots,1]$, the servers store replicated databases. A scheme to retrieve single files from replicated databases for multiple users was proposed in~\cite{XiangZhang}, while guaranteeing server-side privacy. This is different from our setup, even if we  remove the user-side privacy and security conditions, since our robust decoding setup in this case imposes that each user can decode from the signal of any single server (i.e., $L=1$). 
\end{remark}

\begin{remark}[Less Constrained Systems and Naming Convention]\label{rem:LessConstrained}
For any given RSP-LFR $(N,K,L,H)$ system, the robust correctness condition in~\eqref{eqn:correctness} 
guarantees that the users can correctly decode their demands by receiving the signals from \emph{any} $L$ servers.  In addition to investigating the load-memory tradeoff of the RSP-LFR system, we also discuss less constrained systems where some of the conditions in~\eqref{eq:allconditions} are relaxed or dropped. In such systems, the optimal load-memory tradeoff can be similarly defined as in~\eqref{RM:star}.  In particular, we use $R_{\mathcal{C}}^*(M)$ to denote the optimal load-memory tradeoff of a system with only the constrains listed in the label $\mathcal{C}$, which can be any of the following:
\begin{itemize}
\item $\rm{L}$: scalar Linear Function Retrieval (LFR) demands, i.e., the demands $\sd{d}_1,\ldots,\sd{d}_K\in\mathbb{F}_q^{N}$;
\item $\rm{F}$: File Retrieval (FR) demands, i.e., the demands $\sd{d}_1,\ldots,\sd{d}_K$ are restricted to $\{\sd{e}_1,\ldots,\sd{e}_K\}$, where $\sd{e}_n \in \mathbb{F}_q^N, n\in[N],$ is the vector with the $n$-th digit being $1$ and all the others zero;
\item $\rm{S}$: the security condition in~\eqref{eqn:security};
\item $\rm{P}$: both privacy conditions in~\eqref{eqn:user:privacy} and \eqref{eqn:server:privacy}; 
\item $\rm{P}_{\rm{U}}$: the user-side privacy condition in~\eqref{eqn:user:privacy};
\item $\rm{P}_{\rm{S}}$: the server-side privacy condition in~\eqref{eqn:server:privacy};

\end{itemize}
The convention for the subscript $\mathcal{C}$ is:
\begin{enumerate}
\item It contains either $\rm{L}$ or $\rm{F}$, but not both, so as to identify the demand type allowed in the system. 
\item It contains at most one character between $\rm{P},\rm{P}_{\rm{U}},\rm{P}_{\rm{S}},$, which identifies  the privacy condition imposed on the system. 
\item  
The tradeoff is defined for $M\in[1,N]$ if it contains $\rm{S}$, and   
for $M\in[0,N]$ otherwise (see Remark \ref{remark:M}). 
\end{enumerate}
With the above conventions, the value of $\mathcal{C}$ is one from the set
\begin{IEEEeqnarray}{c}
\mathbf{\Omega}:=\{\rm L, LS, LP, LP_S, LP_U, LSP, LSP_S, LSP_U,\notag\\
\rm F, FS, FP, FP_U, FP_S, FSP, FSP_S, FSP_U \}.\IEEEeqnarraynumspace\label{val:C}
\end{IEEEeqnarray}
Notice that, if $\mathcal{C}=\rm{LSP}$, the system is the novel RSP-LFR setup introduced in this paper, thus, $R_{\rm{LSP}}^*(M)=R^*(M)$ in~\eqref{RM:star}, defined for all $M\in[1,N]$. 

We will also need to discuss the single server system where all the files are stored at the server.  The optimal load-memory tradeoff can be similarly defined for such a system for any constraint implied by $\mathcal{C}\in\mathbf{\Omega}$. We will use $\overline{R}_{\mathcal{C}}^*(M)$ to denote the optimal tradeoff in the single server system with constraint identified by $\mathcal{C}\in\mathbf{\Omega}$.  

\end{remark}

\section{PDAs and A Toy Example}\label{sec:PDA:example}
Our achievable results are based on the notion of PDA~\cite{Yan2017PDA}, originally introduced  to reduce the subpacketization in the single-server systems for single file retrieval and without any security or privacy guarantees.
In this section, we first review the definition of PDA, and then give an 
example to highlight the key ideas in the design of our RSP-LFR scheme. 
The general construction will be discussed in the rest of the paper.

\subsection{Placement Delivery Array}

\begin{definition}[PDA~\cite{Yan2017PDA}]\label{def:PDA} For given $K,F\in\mathbb{N}^+$ and $Z,S\in\mathbb{N}$,  an $F\times K$ array
  $\mathbf{A}=[a_{i,j}]$, $i\in [F], j\in[K]$, composed of
  $Z$ specific symbols ``$*$"  in each column and some ordinary symbols $1,\ldots, S$,
  each occurring at least once,  is called a $(K,F,Z,S)$ PDA, if, for
  any two distinct entries $a_{i,j}$ and $a_{i',j'}$,   we have
  $a_{i,j}=a_{i',j'}=s$, for some ordinary symbol $s\in[S]$ only if
  \begin{enumerate}
     \item [a)] $i\ne i'$, $j\ne j'$, i.e., they lie in distinct rows and distinct columns; and
     \item [b)] $a_{i,j'}=a_{i',j}=*$, i.e., the corresponding $2\times 2$  sub-array formed by rows $i,i'$ and columns $j,j'$ must be of the following form
  \begin{IEEEeqnarray}{c}
    \left[\begin{array}{cc}
      s & *\\
      * & s
    \end{array}\right]~\textrm{or}~
    \left[\begin{array}{cc}
      * & s\\
      s & *
    \end{array}\right].
  \end{IEEEeqnarray}
   \end{enumerate}
   \end{definition}

\subsection{A Toy RSP-LFR Example from PDAs}

We derive here a RSP-LFR scheme associated to the $(K,F,Z,S)=(3,3,1,3)$ PDA 
\begin{IEEEeqnarray}{c}
\mathbf{A}=\left[\begin{array}{ccc}
        * & 1 &2\\
        1 & * &3\\
        2& 3&*
      \end{array}
\right]
\label{eqn:pippo}
\end{IEEEeqnarray}
for an $(N,K,L,H)=(4,3,2,3)$ distributed system. 

Let the four files be $W_1,W_2,W_3,W_4\in\mathbb{F}_2^B$ and the $(3,2)$ generator matrix is given by 
\begin{IEEEeqnarray}{c}
G=\left[\begin{array}{ccc}
1&0&1\\
0&1&1
\end{array}
\right].\label{eqn:exam:G}
\end{IEEEeqnarray}
That is, each file is split into $L=2$ subfiles, $W_n=(W_{n,1},W_{n,2}), n\in[4]$ and by~\eqref{eqn:MDS:code}, the contents stored at the servers are
\begin{subequations}
\begin{IEEEeqnarray}{rCcCl}
C_1&=&\overline{W}_{[4],1}&=&W_{[4],1},\\
C_2&=&\overline{W}_{[4],2}&=&W_{[4],2},\\
C_3&=&\overline{W}_{[4],3}&=&W_{[4],1}\oplus W_{[4],2}.
\end{IEEEeqnarray}
\end{subequations}
Based on the PDA $\mathbf{A}$ in~\eqref{eqn:pippo},
each subfile $W_{n,l}$ is partitioned into $F=3$ equal-size  packets, $W_{n,l,1},W_{n,l,2},W_{n,l,3}$ for all $n\in[4],l\in[2]$. Accordingly, the coded subfile $\overline{W}_{n,h}$ is partitioned into $F=3$ equal-size packets, $\overline{W}_{n,h,1},\overline{W}_{n,h,2},\overline{W}_{n,h,3}$.  

Similarly to~\eqref{eqn:LinearFiles}, for any $\sd{a}=(a_1,a_2,a_3,a_4)^\top\in\mathbb{F}_2^4$, we use the following notation to denote the linear combination of (un)coded packets with coefficient vector $\sd{a}$:
\begin{subequations}
\begin{IEEEeqnarray}{rCl}
W_{\sd{a},l,i}&:=&\bigoplus_{n\in[4]}a_n W_{n,l,i}, \\
\overline{W}_{\sd{a},h,i}&:=&\bigoplus_{n\in[4]}a_n\overline{W}_{n,h,i}=\bigoplus_{l\in[2]}g_{l,h}W_{\sd{a},l,i}
 \label{eqn:linear:coded:packet}
\end{IEEEeqnarray}
\end{subequations}
for all $l\in[2],i\in[3],h\in[3]$. 

The system operates as follows.

\emph{Placement Phase:} 
The servers share $LS=6$  vectors $\{V_{l,s}:l\in[2],s\in[3]\}$, which are generated independently and uniformly from $\mathbb{F}_2^{B/6}$, where the packets $V_{1,s},V_{2,s}$ will be associated to the ordinary symbol $s\in[3]$. Each user~$k\in[3]$ generates a random vector $\sd{p}_k=(p_{k,1},p_{k,2},p_{k,3},p_{k,4})^\top\in\mathbb{F}_2^4$.  The cache content of the user~$k$ is composed of $\sd{p}_k$ and the (un)coded packets in the corresponding column in Table \ref{table:Z}.

\begin{table}
\small
\centering
\scalebox{1}{\begin{threeparttable}
\caption{The cached contents of users$^{\dagger}$ according to $\mathbf{A}$ in~\eqref{eqn:pippo}.}\label{table:Z}
\begin{tabular}{ccc}
  \toprule
 User~$1$& User~$2$& User~$3$\\\hline
 $W_{[4],[2],1}$&$W_{\sd{p}_2,[2],1}\oplus V_{[2],1}$&$W_{\sd{p}_3,[2],1}\oplus V_{[2],2}$\\
 $W_{\sd{p}_1,[2],2}\oplus V_{[2],1}$& $W_{[4],[2],2}$& $W_{\sd{p}_3,[2],2}\oplus V_{[2],3}$\\
 $W_{\sd{p}_1,[2],3}\oplus V_{[2],2}$ &$W_{\sd{p}_2,[2],3}\oplus V_{[2],3}$ &$W_{[4],[2],3}$\\
 \bottomrule
\end{tabular}
\begin{tablenotes}
\item[$\dagger$]
\footnotesize{In addition, each user~$k\in[3]$ caches $\sd{p}_k$.}
\end{tablenotes}
\end{threeparttable}}
\end{table}

The packets $W_{[4],[2],i}$ are associated to the $i$-th row of $\mathbf{A}$ in~\eqref{eqn:pippo} and user~$k$ is associated to the $k$-th column of $\mathbf{A}$. The packets in the $i$-th row of Table \ref{table:Z} of  user~$k$ are created according to the entry $a_{i,k}$ of $\mathbf{A}$ in~\eqref{eqn:pippo}: if $a_{i,k}=*$, user~$k$ caches $NL=8$ uncoded packets $W_{[4],[2],i}$, otherwise it caches $L=2$ coded packets $W_{\sd{p}_{k},[2],i}\oplus V_{[2],a_{i,k}}$. 

\emph{Delivery Phase:} 
Assume that user~$1,2,3$ demands the linear combination 
$W_{\sd{d}_1},W_{\sd{d}_2}$ and $W_{\sd{d}_3}$, respectively, where $\sd{d}_1,\sd{d}_2,\sd{d}_3\in\mathbb{F}_2^4$. Each user~$k\in[3]$ sends $\sd{q}_k=\sd{p}_k\oplus \sd{d}_k$ to all the servers as queries. Upon receiving the query vectors $\sd{q}_{[3]}$,  each server $h\in[3]$ sends a signal $X_h$ to the users, where $X_h$ is composed of the query vectors $\sd{q}_{[3]}$ and $S=3$ coded packets as in the Table \ref{table:X}, which are associated to the ordinary symbols $s=1,2,3$ of $\mathbf{A}$ in~\eqref{eqn:pippo}, respectively, where $(\overline{V}_{1,s},\overline{V}_{2,s},\overline{V}_{3,s})$ is the MDS codeword of $(V_{1,s},V_{2,s})$ with generator matrix $G$ in~\eqref{eqn:exam:G}, 
i.e., for $s\in[3]$,
\begin{IEEEeqnarray}{c}
\overline{V}_{1,s}=V_{1,s},~\overline{V}_{2,s}=V_{2,s},~ \overline{V}_{3,s}=V_{1,s}\oplus V_{2,s}.\IEEEeqnarraynumspace 
\end{IEEEeqnarray}

\begin{table*}
\centering
\scalebox{1}{\begin{threeparttable}
\caption{The signals sent by the servers$^\dagger$ according to $\mathbf{A}$ in~\eqref{eqn:pippo}.}\label{table:X}
\begin{tabular}{cccc}
\toprule
$s$& Server $1$&Server $2$&Server $3$\\\hline
$1$&$\overline{V}_{1,1}\oplus \overline{W}_{\sd{q}_1,1,2}\oplus \overline{W}_{\sd{q}_2,1,1}$&$\overline{V}_{2,1}\oplus \overline{W}_{\sd{q}_1,2,2}\oplus\overline{W}_{\sd{q}_2,2,1}$& $\overline{V}_{3,1}\oplus \overline{W}_{\sd{q}_1,3,2}\oplus \overline{W}_{\sd{q}_2,3,1}$\\
 $2$&$\overline{V}_{1,2}\oplus \overline{W}_{\sd{q}_1,1,3}\oplus \overline{W}_{\sd{q}_3,1,1}$&$\overline{V}_{2,2}\oplus \overline{W}_{\sd{q}_1,2,3}\oplus\overline{W}_{\sd{q}_3,2,1}$& $\overline{V}_{3,2}\oplus\overline{W}_{\sd{q}_1,3,3}\oplus \overline{W}_{\sd{q}_3,3,1} $\\
 $3$&$\overline{V}_{1,3}\oplus \overline{W}_{\sd{q}_2,1,3}\oplus\overline{W}_{\sd{q}_3,1,2}$&$\overline{V}_{2,3}\oplus \overline{W}_{\sd{q}_2,2,3}\oplus \overline{W}_{\sd{q}_3,2,2}$& $\overline{V}_{3,3}\oplus \overline{W}_{\sd{q}_2,3,3}\oplus \overline{W}_{\sd{q}_3,3,2} $\\
 \bottomrule
\end{tabular}
\begin{tablenotes}
\item[$\dagger$]
\footnotesize{In addition, each server $h\in[3]$ transmits the query vectors $\sd{q}_{[3]}$.}
\end{tablenotes}
\end{threeparttable}}
\end{table*}

%

\emph{Performance:}
Each user~$k\in[3]$ can decode the linear combination $W_{\sd{d}_k}$ with signals from any $L=2$ servers because user~$k$ can decode $W_{\sd{d}_k,[2],k}$ since it has cached all the uncoded packets $W_{[4],[2],k}$ from Table \ref{table:Z}. For the other packets, we note:

\begin{table}
\centering
\scalebox{1}{
\begin{threeparttable}
\caption{The signals a user can decode from the transmission by the servers$^\dagger$ according to $\mathbf{A}$ in~\eqref{eqn:pippo}.}\label{table:S}
\begin{tabular}{ccc}
\toprule
$s$&Subfile $1$&Subfile $2$\\\hline
$1$&$V_{1,1}\oplus W_{\sd{q}_1,1,2}\oplus W_{\sd{q}_2,1,1}$&$V_{2,1}\oplus W_{\sd{q}_1,2,2}\oplus W_{\sd{q}_2,2,1}$\\
 $2$&$V_{1,2}\oplus W_{\sd{q}_1,1,3}\oplus W_{\sd{q}_3,1,1}$&$V_{2,2}\oplus W_{\sd{q}_1,2,3}\oplus W_{\sd{q}_3,2,1}$\\
 $3$&$V_{1,3}\oplus W_{\sd{q}_2,1,3}\oplus W_{\sd{q}_3,1,2}$&$V_{2,3}\oplus W_{\sd{q}_2,2,3}\oplus W_{\sd{q}_3,2,2}$\\
 \bottomrule
\end{tabular}
\begin{tablenotes}
\item[$\dagger$]
\footnotesize{In addition, each server $h\in[3]$ transmits the query vectors $\sd{q}_{[3]}$.}
\end{tablenotes}
\end{threeparttable}}
\end{table}

\begin{itemize}

\item For each $s\in[3]$, the signals associated to $s$ over the servers form an  MDS codeword with generator matrix $G$, whose original packets are coded packets within each subfile as shown in Table \ref{table:S}. By the property of MDS codes, each user can decode the signals in Table \ref{table:S} by receiving signals from any $L=2$ of the servers.

\item Upon obtaining the signals in Table \ref{table:S}, each user~$k\in[3]$ can proceed with the decoding process for each subfile $l\in[2]$ as in~\cite{Y:D:Privacy}. 
Let us take $s=1$ for subfile $l=1$ as an example. As $a_{1,2}=a_{2,1}=1$, 
user~$1$ can decode $W_{\sd{d}_1,1,2}$ and user~$2$ can decode $W_{\sd{d}_2,1,1}$
from the signal $V_{1,1}\oplus W_{\sd{q}_1,1,2}\oplus W_{\sd{q}_2,1,1}$, i.e.,
\begin{subequations}
\begin{IEEEeqnarray}{rCl}
W_{\sd{d}_1,1,2}&=&(V_{1,1}\oplus W_{\sd{q}_1,1,2}\oplus W_{\sd{q}_2,1,1})\IEEEeqnarraynumspace\\
&&\oplus (V_{1,1}\oplus W_{\sd{p}_1,1,2})\label{exam:decode:b}\\
&&\oplus W_{\sd{q}_2,1,1},\label{exam:decode:c}
\end{IEEEeqnarray}
\end{subequations}
thus, user~$1$ can decode $W_{\sd{d}_1,1,2}$ since the signals in~\eqref{exam:decode:b} are cached by user~$1$, and the signal in~\eqref{exam:decode:c} can be computed from the cached uncoded packets $W_{[4],1,1}$ and the vector $\sd{q}_2$. Similarly, user~$2$ can decode the packet $W_{\sd{d}_2,1,1}$ by computing
\begin{subequations}
\begin{IEEEeqnarray}{rCl}
W_{\sd{d}_2,1,1}&=&(V_{1,1}\oplus W_{\sd{q}_1,1,2}\oplus W_{\sd{q}_2,1,1})\IEEEeqnarraynumspace\\
&&\oplus (V_{1,1}\oplus W_{\sd{p}_2,1,1})\\
&&\oplus W_{\sd{q}_1,1,2}.
\end{IEEEeqnarray}
\end{subequations}

One can verify that each user~$k\in[3]$ can decode all the remaining packets $W_{\sd{d}_k,[2],[3]\backslash\{k\}}$ from its stored contents,  the signals in Table \ref{table:S} and the query vectors $\sd{q}_{[3]}$.      
\end{itemize}
This concludes the proof of correct robust decoding. Privacy and security are guaranteed since each signal is accompanied by a key of random and uniformly distributed bits.

In term of memory-load performance, recall that each packet is of size $\frac{B}{6}$ bits. 
Each user caches $12$ packets and $1$ vectors in $\mathbb{F}_2^4$, whose length does not scale with $B$. 
Thus the needed memory is $M=12\times\frac{1}{6}=2$ files.  Each of the $3$ server sends $3$ packets and $3$ vectors in $\mathbb{F}_2^4$, thus the achieved load is $R=3\times 3\times\frac{1}{6}=\frac{3}{2}$ files. Hence, the scheme achieves the memory-load pair  $(M,R)=\big(2,\frac{3}{2}\big)$.

\section{Main Results}\label{sec:main}
\subsection{PDA based RSP-LFR Schemes}
With any given PDA, we will construct an associated RSP-LFR scheme.
The following theorem summarizes the performance of PDA based SP-LFR scheme, which will be proved by presenting and analyzing the construction in Section~\ref{sec:DSP-LFR:scheme}.
\begin{theorem}\label{thm:PDA} For any $(N,K,L,H)$ system and a given $(K,F,Z,S)$ PDA $\mathbf{A}$, there exists an associated RSP-LFR scheme that achieves the  memory-load pair
\begin{IEEEeqnarray}{c}
\big(M_{\mathbf{A}},R_{\mathbf{A}}\big)=\bigg(1+\frac{Z}{F}(N-1) , \frac{H}{L}\cdot \frac{S}{F} \bigg).\label{eqn:PDA:MR}
\end{IEEEeqnarray}
with subpacketization $LF$.
\end{theorem}

\begin{remark}[Comparison with single-server systems]\label{rem:comparison}
With the procedure described in Section~\ref{sec:DSP-LFR:scheme}, we can easily obtain RSP-LFR schemes from existing PDA constructions, such as those in~\cite{Yan2017PDA,PDA:bipartite,PDA:a,PDA:b,PDA:c}. 
If $H=L=1$, the system degrades to a single-server shared-link system, where all the files are stored at the server~\cite{Maddah-Ali2014fundamental}. In~\cite{Y:D:SP-LFR}, a key superposition scheme was proposed to guarantee the correctness, security, and user privacy conditions simultaneously based on any  $(K,F,Z,S)$ PDA $\mathbf{A}$ for single-server systems.  The scheme in~\cite{Y:D:SP-LFR} achieves the memory-load pair in~\eqref{eqn:PDA:MR} with $H/L=1$. In other words, the RSP-LFR scheme with PDA $\mathbf{A}$ achieves the same memory size as in the single server case but the load is scaled by a factor $\frac{H}{L}$.
In the case $H=L$, each user needs to retrieve information from all the servers, and the total load is the same as that from a single server case (i.e., $H=L=1$). 
Moreover, this indicates that, in addition to guaranteeing correctness, security, and user-side privacy conditions,  the server-side privacy condition does not increase the load-memory tradeoff  in the non-robust multi-server case with $H=L$. 
\end{remark}

\subsection{Optimality of MAN-PDA based RSP-LFR Schemes}

The following PDA describing the MAN scheme in~\cite{Maddah-Ali2014fundamental} 
is important, and will be referred to as MAN-PDA in the following.

\begin{definition}[MAN-PDA]\label{def:MNPDA}
For any integer  $j\in[0:K]$, define the set
$\mathbf{\Omega}_j\triangleq\{\mathcal{T}\subseteq[K]:|\mathcal{T}|=j\}$.
Fix any integer $t\in[0:K]$, denote the set  $\mathbf{\Omega}_t=\{\mathcal{T}_i : i\in[ {K \choose t}] \}$. Also, choose an arbitrary bijective function $\kappa_{t+1}$ from $\mathbf{\Omega}_{t+1}$
to the set $\big[{K \choose {t+1}}\big]$.
Then, define the array $\mathbf{A}_t=[a_{i,j}]$ as
	\begin{IEEEeqnarray}{c}
		a_{i,j}\triangleq \left\{\begin{array}{ll}
			*, &\textnormal{if}~j\in\mathcal{T}_i \\
			\kappa_{t+1}(\{j\} \cup \mathcal{T}_{i}), &\textnormal{if}~j\notin\mathcal{T}_i
		\end{array}
		\right..
		\label{eq:MAN:aij}
	\end{IEEEeqnarray}
\end{definition}

It was proved in~\cite{Yan2017PDA} that $\mathbf{A}_t$ from~\eqref{eq:MAN:aij} 
in Definition~\ref{def:MNPDA} is a $(K,{K\choose t},{K-1\choose t-1},{K\choose t+1})$ PDA.

\begin{example}[A MAN-PDA]
Consider $K=4$, $t=2$, let $\mathcal{T}_1=\{1,2\},\mathcal{T}_2=\{1,3\},\mathcal{T}_3=\{1,4\},\mathcal{T}_4=\{2,3\},\mathcal{T}_5=\{2,4\}$ and $\mathcal{T}_6=\{3,4\}$. Let $\kappa_3$ be the lexicographic order of a subset of size  $3$ in  $\mathbf{\Omega}_3$, e.g., $\kappa_3(\{1,2,3\})=1,\kappa_3(\{1,2,4\})=2$ and $\kappa_3(\{1,3,4\})=3$ and $\kappa_3(\{2,3,4\})=4$.
The corresponding $(4,6,3,4)$ PDA is given by 
\begin{IEEEeqnarray}{c}
\mathbf{A}_2=\left[\begin{array}{cccc}
*&*&1&2\\
*&1&*&3\\
*&2&3&*\\
1&*&*&4\\
2&*&4&*\\
3&4&*&*
\end{array}
\right].
\end{IEEEeqnarray}
\end{example}

The following theorem summarizes the performance of MAN-PDA and its optimality. The proof is presented in Section~\ref{sec:MAN:PDA}
\begin{theorem}\label{thm:MAN}
Let $R(M)$ be the 
lower convex envelope of the following points
\begin{IEEEeqnarray}{rCl}
(M_t, R_t)= \left(1+\frac{t(N-1)}{K}, \frac{H(K-t)}{L(t+1)}\right),\IEEEeqnarraynumspace\label{eqn:MRt}
\end{IEEEeqnarray}
where $t\in[0:K]$,
then $R(M)$ is achievable in an $(N,K,L,H)$ RSP-LFR system, where the point $(M_t,R_t)$ can be achieved with subpacketization  $L{K\choose t}$. Moreover, $R(M)$ and the optimal communication load $R^*(M)$ satisfies
\begin{enumerate}
  \item $N\geq K$, for all $M\in[1,N)$,
  \begin{IEEEeqnarray}{c}
  \frac{R(M)}{R^*(M)}\leq \left\{\begin{array}{ll}
1,&\textnormal{if}~K=1\\
                                                    2, &\textnormal{if}~N=K=2  \\
                                                    6.02652, &\textnormal{if}~N=K\geq 3  \\
                                                    5.0221, & \textnormal{if}~N=K+1 \\
                                                    4.01768, &\textnormal{if}~N\geq K+2
                                                  \end{array}
  \right..\IEEEeqnarraynumspace\label{eqn:gap1}
  \end{IEEEeqnarray}
  \item $N<K$, for all $M\in[2,N)$,
  \begin{IEEEeqnarray}{c}
  \frac{R(M)}{R^*(M)}<8.\label{eqn:gap2}
  \end{IEEEeqnarray}
\end{enumerate}
\end{theorem}

\begin{remark}[Open regime $N<K,1\leq M<2$] In the regime $N<K,1\leq M<2$ the gap is unbounded. From our proof, $\frac{R(M)}{R^*(M)}$ is upper bounded by $\frac{\overline{R}_{\rm{LSP_U}}(M)}{\overline{R}_{\rm{LSP_U}}^*(M)}$, where $\overline{R}_{\rm{LSP_U}}(M)$ is the tradeoff achieved by the key superposition scheme in the single server system where the security and user-side privacy conditions are imposed \cite{Y:D:SP-LFR}, and $\overline{R}_{\rm{LSP_U}}^*(M)$ is the corresponding optimal tradeoff. The gap result in Theorem \ref{thm:MAN} thus follows from the bound for $\frac{\overline{R}_{\rm{LSP_U}}(M)}{\overline{R}_{\rm{LSP_U}}^*(M)}$ in \cite{Y:D:SP-LFR}, where the same regime is open. The main problem in this regime for the single server model is that, if security keys are used`\cite{Y:D:SP-LFR, Security}, for the point $M=1$ the best know achievable load is $K$, while the best known converse is $N$. When new converse and gap will be obtained for this regime in the single server case, the same gap will apply to our RSP-LFR system.  
\end{remark}

The following theorem implies that, with the given procedure of deriving RSP-LFR scheme in Section~\ref{sec:DSP-LFR:scheme}, the memory-load pairs $\{(M_t,R_t):t\in[0:K]\}$ achieved by the MAN-PDAs are Pareto-optimal among all PDA based RSP-LFR schemes. Moreover, the MAN-PDAs have the smallest subpacketization among all PDA based RSP-LFR schemes achieving these points.  The proof is deferred to Section~\ref{sec:optimal:subpacket}.
\begin{theorem}\label{thm:MAN:PDA} Given a $(K,F,Z,S)$ PDA, 
if the associated RSP-LFR scheme achieves a memory-load pair $(M,R)$,
then necessarily
\begin{IEEEeqnarray}{c}
R\geq\frac{HK(N-M)}{L(N-1+K(M-1))}=\left.\frac{H(K-x)}{L(x+1)}\right|_{x = K\frac{M-1}{N-1}}.\label{eqn:PDA:bd}
\end{IEEEeqnarray}
In particular, the memory-load pairs $\{(M_t,R_t):t\in[0:K]\}$ satisfy~\eqref{eqn:PDA:bd} with  equality. Moreover, if $M=M_t$ and $R=R_t$ for some $t\in[0:K]$, then the subpacketization is at least  $L{K\choose t}$.
\end{theorem}

\begin{remark}[Subpacketizations]
By the procedure described in Section~\ref{sec:DSP-LFR:scheme}, we can easily obtain RSP-LFR schemes from existing PDA constructions, such as those in~\cite{Yan2017PDA,PDA:bipartite,PDA:a,PDA:b,PDA:c}.  
It was showed in~\cite{Y:D:SP-LFR} that the PDA-based construction in~\cite{Yan2017PDA} achieves a slightly larger load than MAN-PDA for the same memory size, while reducing the subpacketization by a factor that increases exponentially with $K$. Thus, PDAs in~\cite{Yan2017PDA} sacrifice some load for an exponential reduction in subpacketization.
\end{remark}

\subsection{Improved Load-Memory Tradeoffs Less Constrained Systems}\label{subsec:degraded}
Obviously, the load-memory tradeoff $R(M)$ in Theorem \ref{thm:MAN} is achievable for any less constrained system described in Remark~\ref{rem:LessConstrained}.   
In this subsection, we present improved achievable results for the following three less constrained systems.
The details are presented in Section~\ref{sec:degraded}.

\subsubsection{Robust Private Linear Function Retrieval (RP-LFR) System ($\mathcal{C}=\rm{LP}$)} 
In an $(N,K,L,H)$ RP-LFR system, 
the correctness condition~\eqref{eqn:correctness}  and the privacy conditions~\eqref{eqn:user:privacy}--\eqref{eqn:server:privacy} must be guaranteed for all LFR demands. 

\begin{theorem}\label{thm:RP-LFR}
For an $(N,K,L,H)$ RP-LFR system, let  $R_{\rm{LP}}(M)$ be the lower convex envelope of the point $\big(0,\frac{HN}{L}\big)$ and the following points 
\begin{IEEEeqnarray}{rl}
&\lefteqn{(M_t^{\rm{LP}},R_t^{\rm{LP}}):=}\notag\\
&\bigg(1+\frac{t(N-1)}{K},\frac{H\big({K\choose t+1}-{K-\min\{K,N\}\choose t+1}\big)}{L{K\choose t}}\bigg),\IEEEeqnarraynumspace\label{eqn:MR:RPL}
\end{IEEEeqnarray}
where $t\in[0:K]$.
Then,  $R_{\rm{LP}}(M)$ is achievable, and it satisfies 
\begin{IEEEeqnarray}{c}
\frac{R_{\rm{LP}}(M)}{R^*_{\rm{LP}}(M)}\leq 6.3707, \quad\forall\, M\in[0,N].
\end{IEEEeqnarray}
\end{theorem}

\subsubsection{Robust Private File Retrival (RP-FR) System ($\mathcal{C}=\rm{FP}$)} 
In an $(N,K,L,H)$ RP-FR system, 
the correctness condition~\eqref{eqn:correctness}  and the privacy conditions~\eqref{eqn:user:privacy}--\eqref{eqn:server:privacy} must be guaranteed for all FR demands.
 
\begin{theorem}\label{thm:RP-FR}
For an $(N,K,L,H)$ RP-FR system, let  $R_{\rm{FP}}(M)$ be the lower convex envelope of the point $\big(0,\frac{HN}{L}\big)$ and the following points 
\begin{IEEEeqnarray}{rl}
&\lefteqn{(M_t^{\rm{FP}},R_t^{\rm{FP}}):=}\notag\\
&\bigg(1+\frac{t(N-1)}{K},\frac{H\big({K\choose t+1}-{K-\min\{K,N-1\}\choose t+1}\big)}{L{K\choose t}}\bigg),\IEEEeqnarraynumspace
\end{IEEEeqnarray}
where $t\in[0:K]$.
Then,  $R^{\rm{RP\textnormal{-}F}}(M)$ is achievable, and it satisfies
\begin{IEEEeqnarray}{c}
\frac{R_{\rm{FP}}(M)}{R^*_{\rm{FP}}(M)}\leq 5.4606,\quad \forall\, M\in[0,N].
\end{IEEEeqnarray}
\end{theorem}

\subsubsection{Robust Linear Function Retrieval (R-LFR) System ($\mathcal{C}=\rm{L}$)} 
In an $(N,K,L,H)$ R-LFR system, 
only the correctness condition~\eqref{eqn:correctness} must be guaranteed for all LFR demands. 

\begin{theorem}\label{thm:R-LFR}
For an $(N,K,L,H)$ R-LFR system, let  $R_{\rm{L}}(M)$ be the lower convex envelope of the following points 
\begin{IEEEeqnarray}{c}
(M_t^{\rm{L}},R_t^{\rm{L}}):=\bigg(\frac{tN}{K},\frac{H\big({K\choose t+1}-{K-\min\{K,N\}\choose t+1}\big)}{L{K\choose t}}\bigg),\IEEEeqnarraynumspace
\end{IEEEeqnarray}
where $t\in[0:K]$.
Then,  $R_{\rm{L}}(M)$ is achievable and it satisfies
\begin{IEEEeqnarray}{c}
\frac{R_{\rm{L}}(M)}{R^*_{\rm{L}}(M)}\leq 2.00884,\quad\forall\, M\in[0,N].
\end{IEEEeqnarray}
\end{theorem}

\begin{remark}[Less Constrained Systems] 
Notice that if $R_{\mathcal{C}}(M)$ is achievable for the constraint $\mathcal{C}$, then $R_{\mathcal{C}}(M)$ is achievable for all constrains that are less restrictive than $\mathcal{C}$. In particular, with Theorem~\ref{thm:MAN}, the tradeoff $R(M)$ is achievable for all $\mathcal{C}\in\mathbf{\Omega}$. Moreover, with Theorems~\ref{thm:RP-LFR}--\ref{thm:R-LFR}, the tradeoff
\begin{enumerate}
\item $R_{\rm{LP}}(M)$ is achievable for any $\mathcal{C}\in\{\rm LP, LP_S, LP_U\}$;

\item $R_{\rm{FP}}(M)$ is achievable for any $\mathcal{C}\in\{\rm FP, FP_S, FP_U\}$;
 
\item $R_{\rm{L}}(M)$ is achievable for any $\mathcal{C}\in\{\rm{L},\rm{F}\}$.
\end{enumerate}
Moreover, from the proofs in Section \ref{sec:degraded}, it is clear that the subpacketzation for $(M_t^{\mathcal{C}},R_t^{\mathcal{C}})$ is $L{K\choose t}$ for all $t\in[0:K]$ and $\mathcal{C}\in\{\rm LP, FP, L\}$ (and thus also for their less constrained systems).
\end{remark}

\section{Proof of Theorem~\ref{thm:PDA}} 
\label{sec:DSP-LFR:scheme}
In this section, we derive a RSP-LFR scheme for an $(N,K,L,H)$ system from any given $(K,F,Z,S)$ PDA $\mathbf{A}=[a_{i,j}]_{F\times K}$. 
Based on  $\mathbf{A}$, each subfile $W_{n,l}$ ($n\in[N],l\in[F]$)  is partitioned into $F$ equal-size packets, denoted by $W_{n,l,1},\ldots,W_{n,l,F}$, where each packet $W_{n,l,i}\in\mathbb{F}_q^{B/(LF)}$. 
The packets with index $i$,  i.e., $W_{[N],[L],i}$,  are associated to the $i$-th row of $\mathbf{A}$. According to~\eqref{eqn:MDS:code}, each coded subfile $\overline{W}_{n,h}$ ($n\in[N],h\in[H]$) is composed of $F$ coded packets, denoted by $\overline{W}_{n,h,1},\ldots,\overline{W}_{n,h,F}$, where
\begin{IEEEeqnarray}{rCl}
\overline{W}_{n,h,i}&=&\sum_{l\in[L]}g_{l,h}W_{n,l,i},\quad\forall\,i\in[F].
\end{IEEEeqnarray}
That is, the coded contents stored at server $h$ 
are
\begin{IEEEeqnarray}{c}
C_h=\overline{W}_{[N],h,[F]},\quad\forall\, h\in[H].
\end{IEEEeqnarray}

We use the following notations similarly to~\eqref{eqn:LinearFiles} for any $\sd{a}=(a_1,\ldots,a_N)^{\top}\in\mathbb{F}_q^{N}$ to denote the linear combination of (un)coded packets:
\begin{IEEEeqnarray}{rCll}
W_{\sd{a},l,i}&=&\sum_{n\in[N]}a_n W_{n,l,i},\quad&\forall\,l\in[L],i\in[F].\IEEEeqnarraynumspace\\
\overline{W}_{\sd{a},h,i}&=&\sum_{n\in[N]}a_n\overline{W}_{n,h,i},\quad &\forall\, h\in[H],i\in[F].\IEEEeqnarraynumspace
\end{IEEEeqnarray}
Notice that   $(\overline{W}_{\sd{a},1,i},\ldots,\overline{W}_{\sd{a},H,i})$ is the MDS codeword of $(W_{\sd{a},1,i},\ldots,W_{\sd{a},L,i})$, i.e.,
\begin{IEEEeqnarray}{rl}
&\lefteqn{(\overline{W}_{\sd{a},1,i},\ldots,\overline{W}_{\sd{a},H,i})}\notag\\
&=\Big(\sum_{l\in[L]}g_{l,1}W_{\sd{a},l,i},\ldots,\sum_{l\in[L]}g_{l,H}W_{\sd{a},l,i}\Big).\label{eqn:MDS:packets}
\end{IEEEeqnarray}
Moreover, $W_{\sd{a},l,i},\overline{W}_{\sd{a},h,i}$ are linear in $\sd{a}$. 

\paragraph*{Placement Phase} the servers share the random variables
\begin{IEEEeqnarray}{c}
V=\{V_{l,s}:l\in[L],s\in[S]\},\label{eqn:random:V}
\end{IEEEeqnarray}
which are $SL$ vectors  independently and uniformly distributed over $\mathbb{F}_q^{B/(FL)}$. Each user~$k\in[K]$ locally generates a random vector $\sd{p}_k$ uniformaly over $\mathbb{F}_q^N$, and constructs its local cache $Z_k$ as
\begin{subequations}\label{eqn:Zk}
\begin{IEEEeqnarray}{rCl}
&&Z_k=\{\sd{p}_k\}\label{eqn:Zk:c}
\\
&&\bigcup\{W_{n,l,i}:n\in[N],l\in[L], i\in[F],a_{i,k}=*\}\IEEEeqnarraynumspace\label{eqn:Zk:a}\\
&&\bigcup\{W_{\sd{p}_k,l,i}+V_{l,a_{i,k}}:l\in[L],i\in[F],a_{i,k}\neq *\}.\IEEEeqnarraynumspace\label{eqn:Zk:b}
\end{IEEEeqnarray}
\end{subequations}

\paragraph*{Delivery Phase} Assume that user~$k\in[K]$ demands 
$W_{\sd{d}_k}$, for some $\sd{d}_k\in\mathbb{F}_q^N$.  Then user~$k\in[K]$ sends query $\sd{q}_k=\sd{d}_k+\sd{p}_k$ to all the servers, i.e., the queries $Q_{k,[H]}$ are constructed as 
\begin{IEEEeqnarray}{c}
Q_{k,h}=\sd{q}_k=\sd{d}_k+\sd{p}_k,\quad \forall\,h\in[H].\label{eqn:Qkh}
\end{IEEEeqnarray} 
For each $s\in[S]$, consider the MDS coded version of $(V_{1,s},\ldots,V_{L,s})$ with the generator matrix $G$, i.e., 
\begin{IEEEeqnarray}{c}
(\overline{V}_{1,s},\ldots,\overline{V}_{H,s})
=\Big(\sum_{l\in[L]}g_{l,1}V_{l,s},\ldots,\sum_{l\in[L]}g_{l,H}V_{l,s}\Big).\IEEEeqnarraynumspace\label{eqn:MDS:V}
\end{IEEEeqnarray}
Upon receiving the queries $Q_{[K],h}=\sd{q}_{[K]}$, each server $h\in[H]$ sends the the signal
\begin{IEEEeqnarray}{c}
X_h=(\sd{q}_{[K]},\overline{Y}_{h,[S]})\label{eq:Xh}
\end{IEEEeqnarray}
to the users, where for each $s\in[S]$, $\overline{Y}_{h,s}$ is 
\begin{IEEEeqnarray}{c}
\overline{Y}_{h,s}=\overline{V}_{h,s}+\sum_{\substack{(u,v)\in[F]\times[K]\\a_{u,v}=s}}\overline{W}_{\sd{q}_v,h,u}.\label{eqn:Yhs}
\end{IEEEeqnarray}


\paragraph*{Robust Correctness}
We need to show that for each user~$k\in[K]$,  with any $\mathcal{L}\subseteq[K]$ such that $|\mathcal{L}|=L$, user~$k$ can decode its demanded scalar linear function $W_{\sd{d}_k}$, i.e., all the packets $W_{\sd{d}_k,[L],[F]}$.

For each $i\in[F]$ such that $a_{i,k}=*$, by~\eqref{eqn:Zk:a}, user~$k\in[K]$ has stored all the packets $W_{[N],[L],i}$, thus it can directly compute the packets $W_{\sd{d}_k,l,i}$ for each $l\in[L]$. 

Now, consider any $i\in[F]$ such that $a_{i,k}\neq *$. Let $s\triangleq a_{i,k}$, notice that by~\eqref{eqn:MDS:packets} and~\eqref{eqn:MDS:V}, 
$(\overline{Y}_{1,s},\ldots,\overline{Y}_{H,s})$ is  the MDS coded version of information coded packets  $(Y_{1,s},\ldots,Y_{L,s})$ with generator matrix $G$, where
\begin{IEEEeqnarray}{c}
Y_{l,s}:=V_{l,s}+\sum_{\substack{(u,v)\in[F]\times[K]\\a_{u,v}=s}}W_{\sd{q}_v,l,u},\ \forall\, l\in[L].\label{eqn:information:Yls}
\end{IEEEeqnarray}
By the property of MDS codes, each user can decode all the $L$ coded packets in~\eqref{eqn:information:Yls} with signals from any subset of $L$ servers for each $s\in[S]$. 
Since $a_{i,k}=s$, for each $l\in[L]$, the signal $Y_{l,s}$ in~\eqref{eqn:information:Yls} can be written as
\begin{IEEEeqnarray}{rCl}
Y_{l,s}
&=&V_{l,s}+W_{\sd{q}_k,l,i}+\sum_{\substack{(u,v)\in[F]\times[K]\\a_{u,v}=s,(u,v)\neq (i,k)}}W_{\sd{q}_v,l,u}\IEEEeqnarraynumspace\\
&\overset{(a)}{=}&W_{\sd{d}_k,l,i}+(V_{l,a_{i,k}}+W_{\sd{p}_k,l,i})\notag\\
&&\quad\quad +\sum_{\substack{(u,v)\in[F]\times[K]\\a_{u,v}=s=a_{i,k},(u,v)\neq(i, k)}}W_{\sd{q}_v,l,u},\label{eqn:Yhs:analysis}
\end{IEEEeqnarray}
where $(a)$ follows from $\sd{q}_k=\sd{p}_k+\sd{d}_k$. Therefore, user~$k\in[K]$ can decode $W_{\sd{d}_k,l,i}$ from the the signal $Y_{l,s}$ by canceling the remaining terms since 
\begin{enumerate}
\item the coded packet $V_{l,a_{i,k}}+W_{\sd{p}_k,l,i}$ is cached by user~$k$ by~\eqref{eqn:Zk:b};
\item for each $(u,v)\in[F]\times[K]$ such that $a_{u,v}=s$ and $(u,v)\neq (i,k)$, since $a_{i,k}=a_{u,v}=s$, by the definition of PDA,  $i\neq u,v\neq k$ and $a_{i,v}=a_{u,k}=*$. Thus, user~$k\in[K]$ stores all the packets $W_{[N],[L],u}$. Hence, user~$k$ can compute $W_{\sd{q}_v,l,u}$ for each $l\in[L]$.  
\end{enumerate}

\begin{remark}[On the Robust Decoding]
From the above decoding process, user~$k\in[K]$ can decode its demanded linear function if for any $i\in[F]$ such that $a_{i,k}\neq *$, user~$k$ can receive any $L$ of the coded signals $\overline{Y}_{1,a_{i,k}},\ldots,\overline{Y}_{H,a_{i,k}}$. This is less restrictive than the assumptions in our setup (i.e., each user can obtain a fixed subset of signals $X_{\mathcal{L}}$), since (i) it allows the available subset $\mathcal{L}$ to vary over different transmission $s\in[S]$; (ii) it only needs to decode packets over the signals assocatated to $s$ such that, $a_{i,k}=s$ for some $i\in[F]$, which indicates that for $s\in[S]\backslash\{a_{i,k}:i\in[F]\}$, the availability of the signals $\overline{Y}_{[H],s}$ does not affect the decodability of user~$k$. 

\end{remark}

\paragraph*{Security}
We have
\begin{subequations}
\begin{IEEEeqnarray}{rCl}
&&I(W_{[N]};X_{[H]})\label{eqn:security:exp:1}\\
&=&I(W_{[N]};\sd{q}_{[K]},\overline{Y}_{[H],[S]})\\
&=&I(W_{[N]};\sd{q}_{[K]},Y_{[L],[S]})\label{eqn:security:exp:3}\\
&=&I(W_{[N]};\sd{q}_{[K]})+I(W_{[N]};Y_{[L],[S]}\,|\,\sd{q}_{[K]})\IEEEeqnarraynumspace\\
&=&0,\label{eq:exp5}
\end{IEEEeqnarray}
\end{subequations}
where: 
\eqref{eqn:security:exp:3} holds since  $(\overline{Y}_{1,s},\ldots,\overline{Y}_{H,s})$ is the  MDS coded version of $(Y_{1,s},\ldots,Y_{L,s})$ for each $s\in[S]$, and hence  they determine each other; and
\eqref{eq:exp5} follows since (a) the vectors $\sd{q}_{[K]}=\sd{d}_{[K]} + \sd{p}_{[K]}$  are independent of $W_{[N]}$, and
(b) $Y_{[L],[S]}$ are independent of $(W_{[N]},\sd{q}_{[K]})$ because the random variables $V_{[L],[S]}$ are independently and uniformly distributed. 

\paragraph*{User-side Privacy in~\eqref{eqn:user:privacy}} 
We have
\begin{subequations}\label{eqn:user:privacy:proof}
\begin{IEEEeqnarray}{rCl}
&&I(\sd{d}_{[K]\backslash\mathcal{S}};Z_{\mathcal{S}},X_{[H]},\sd{d}_{\mathcal{S}}\,|\,W_{[N]})\\
&=&I(\sd{d}_{[K]\backslash\mathcal{S}};Z_{\mathcal{S}},\sd{q}_{[K]},\overline{Y}_{[H],[S]},\sd{d}_{\mathcal{S}}\,|\,W_{[N]})\IEEEeqnarraynumspace\\
&=&I(\sd{d}_{[K]\backslash\mathcal{S}};Z_{\mathcal{S}},\sd{q}_{[K]},Y_{[L],[S]},\sd{d}_{\mathcal{S}}\,|\,W_{[N]})\IEEEeqnarraynumspace\label{eqn:exp:userP1}\\
&=&0,\label{eqn:exp:userP2}
\end{IEEEeqnarray}
\end{subequations}
where:
\eqref{eqn:exp:userP1} follows since $\overline{Y}_{[H],[S]}$ and $Y_{[L],[S]}$ determine each other due to the fact that $\overline{Y}_{[H],s}$ is the MDS coded version of $Y_{[L],s}$ for each $s\in[S]$; and 
\eqref{eqn:exp:userP2} follows since $\sd{d}_{[K]\backslash\mathcal{S}}=\sd{q}_{[K]\backslash\mathcal{S}}-\sd{p}_{[K]\backslash\mathcal{S}}$ is independent of $(Z_{\mathcal{S}},\sys{W}_{[N]}, \sd{q}_{[K]}, \sd{d}_\mathcal{S},Y_{[L],[H]})$ since $\sd{p}_{[K]\backslash\mathcal{S}}$ are independently and uniformly distributed.

\paragraph*{Server-side Privacy  in~\eqref{eqn:server:privacy}} We have
\begin{subequations}\label{eqn:server:privacy:proof}
\begin{IEEEeqnarray}{rCl}
&&I(\sd{d}_{[K]};Q_{[K],[H]}\,\overline{W}_{[N],[H]},V)\\
&=&I(\sd{d}_{[K]}; \sd{q}_{[K]},W_{[N]},V)\label{eq:server:1}\\
&=&I(\sd{d}_{[K]}; W_{[N]},V)+I(\sd{d}_{[K]}; \sd{q}_{[K]}\,|\,W_{[N]},V)\IEEEeqnarraynumspace\\
&=&0,\label{eqn:exp:ser-privacy}
\end{IEEEeqnarray}
\end{subequations}
where:
\eqref{eq:server:1} follows from~\eqref{eqn:Qkh} and the fact  $\overline{W}_{[N],[H]}$ and $W_{[N]}$ determines each other; and
\eqref{eqn:exp:ser-privacy} holds because (a)  $\sd{d}_{[K]}$ is independent of  $(W_{[N]},V) $; (b) $\sd{q}_{[K]}=\sd{p}_{[K]}+\sd{d}_{[K]}$  are independent of $(\sd{d}_{[K]},W_{[N]},V) $ since the vectors $\sd{p}_{[K]}$ are independent random variables uniformly distributed. 


\paragraph*{Performance}
By~construction, each subfile is split into $F$ equal-size packets, each of length $\frac{B}{LF}$ symbols, thus the subpacketization is $LF$. For each user~$k\in[K]$, by the cached content in~\eqref{eqn:Zk}, for each $i\in[F]$ such that $a_{i,k}=*$, there are $LN$ associated packets cached by the user, one from each file (see~\eqref{eqn:Zk:a}). For each $i\in[F]$ such that $a_{i,k}\neq *$, there are $L$ associated coded packet cached at the user (see~\eqref{eqn:Zk:b}). In addition, the $\sd{p}_{k}$ in~\eqref{eqn:Zk:c} can be stored with $N$ symbols.
 Recall that, each column of a $(K,F,Z,S)$ PDA has $Z$ $``*"$s and $F-Z$ ordinary symbols, thus, the needed cache size is
\begin{IEEEeqnarray}{rCl}
M_\mathbf{A}&=&\inf_{B\in\mathbb{N}^+}\frac{1}{B}\Big((Z \ LN+(F-Z)L)\frac{B}{LF}+N\Big)\IEEEeqnarraynumspace\\
&=&\frac{F+Z \ (N-1)}{F}.\IEEEeqnarraynumspace
\end{IEEEeqnarray}
By~\eqref{eq:Xh}, each server $h\in[H]$ sends $S$ coded packets $Y_{h,[S]}$, each of size $\frac{B}{LF}$ symbols, and the coefficient vectors  $\sd{q}_{[K]}$ can be sent in $KN$ symbols, thus the achieved load is
\begin{IEEEeqnarray}{c}
R_{\mathbf{A}}=\inf_{B\in\mathbb{N}^+}\frac{1}{B}\Big(HS \ \frac{ B}{LF}+H \ KN\Big)=\frac{HS}{LF}.\IEEEeqnarraynumspace
\end{IEEEeqnarray}

\section{MAN-PDA and Its Optimality}
\subsection{MAN-PDA: Performance and Gap Results (Proof of Theorem~\ref{thm:MAN})}\label{sec:MAN:PDA}
The achievability of the point $(M_t,R_t)$ directly follows from Theorem~\ref{thm:PDA} and the $(K,{K\choose t},{K-1\choose t-1},{K\choose t+1})$ MAN-PDA $\mathbf{A}_t$ in Definition~\ref{def:MNPDA}. 
Moreover, the lower convex envelope of the points in~\eqref{eqn:MRt} can be achieved by memory-sharing technique~\cite{Maddah-Ali2014fundamental}.

For the gap result, we derive the following lemma for any $\mathcal{C}\in\mathbf{\Omega}$.
\begin{lemma}\label{lemma:bound} For any $\mathcal{C}\in\mathbf{\Omega}$, for any feasible\footnote{If $\mathcal{C}$ contains $\rm{S}$, $M\in[1,N]$;  else $M\in[0,N]$ (see Remark \ref{remark:M}).} $M$,
\begin{IEEEeqnarray}{c}
R_{\mathcal{C}}^*(M)\geq \frac{H}{L}\cdot \overline{R}_{\mathcal{C}}^*(M),
\end{IEEEeqnarray}
\end{lemma} 
\begin{IEEEproof} For a $(N,K,L,H)$ system with the constraint $\mathcal{C}$, for any feasible design of caches $Z_{[K]}$ and signals $X_{[H]}$ satisfying the constraint $\mathcal{C}$,  for any $\mathcal{L}\subseteq[H]$, the contents $Z_{[K]}$ and signal $X\triangleq X_{\mathcal{L}}$ are a feasible scheme for the single server system with the same constraint $\mathcal{C}$. Thus,
\begin{IEEEeqnarray}{c}
\frac{H(X_{\mathcal{L}})}{B}\geq \overline{R}_{\mathcal{C}}^*(M),\quad \forall\,\mathcal{L}\subseteq[K],\,|\mathcal{L}|=L.\IEEEeqnarraynumspace
\end{IEEEeqnarray}
Therefore, 
\begin{subequations}
\begin{IEEEeqnarray}{rCl}
R_{\mathcal{C}}^*(M)&\geq& \frac{1}{B}\sum_{h\in[H]}H(X_h)\\
&=&\frac{H}{B}\cdot\frac{1}{H}\sum_{h\in[H]}H(X_h)\\
&\geq&\frac{H}{B}\cdot\frac{1}{{H\choose L}}\sum_{\mathcal{L}\subseteq[H],|\mathcal{L}|=L}\frac{H(X_{\mathcal{L}})}{L}\label{eqn:Han}\\
&=&H\cdot\frac{1}{{H\choose L}}\sum_{\mathcal{L}\subseteq[H],|\mathcal{L}|=L}\frac{R_{\mathcal{C}}^*(M)}{L}\label{eqn:abovebound}\\
&\geq &\frac{H}{L}\cdot \overline{R}_{\mathcal{C}}^*(M),
\label{eqn:bound:R:star}
\end{IEEEeqnarray}
\end{subequations}
where~\eqref{eqn:Han} follows from Han's inequality~\cite{Inform_Cover_book}. 
\end{IEEEproof}

Let $\overline{R}_{\rm{LSP_U}}(M)$ be  the lower convex envelope of the following points: for each $t\in[0:K]$,
\begin{IEEEeqnarray}{c}
\big(\overline{M}_{t},\overline{R}_{t}\big)=\Big(1+\frac{t(N-1)}{K},\frac{K-t}{t+1}\Big),\label{eqn:MRt:single}
\end{IEEEeqnarray}
Notice that $\overline{R}_{\rm{LSP_U}}(M)$ is achievable by the key superposition scheme in~\cite{Y:D:SP-LFR} for the single server system with constraint  $\rm{LSP_U}$. 
Comparing~\eqref{eqn:MRt} with~\eqref{eqn:MRt:single}, we see that $R(M)=\frac{H}{L}\cdot \overline{R}_{\rm{LSP_U}}(M)$ 
(see also Remark~\ref{rem:comparison}), 
hence by Lemma \ref{lemma:bound}, for all $M\in[1,N]$,
\begin{IEEEeqnarray}{c}
\frac{R(M)}{R^*(M)}\leq \frac{\overline{R}_{\rm{LSP_U}}(M)}{\overline{R}_{\rm{LSP}}^*(M)}\overset{(a)}{\leq} \frac{\overline{R}_{\rm{LSP_U}}(M)}{\overline{R}_{\rm{LSP_U}}^*(M)}.\label{eqn:gap}
\end{IEEEeqnarray}
where $(a)$ follows from the fact $\overline{R}_{\rm{LSP}}^*(M)\geq\overline{R}_{\rm{LSP_U}}(M)$, since the constraint $\rm{LSP}$ is stronger than the constraint $\rm{LSP_U}$. Thus, the claimed multiplicative gap result  directly follows from~\eqref{eqn:gap} and the bound for $\frac{\overline{R}_{\rm{LSP_U}}(M)}{\overline{R}_{\rm{LSP_U}}^*(M)}$ in \cite[Theorem 3]{Y:D:SP-LFR}.

\subsection{MAN-PDA:Optimality within PDA Based RSP-LFR Schemes (Proof of Theorem~\ref{thm:MAN:PDA})}\label{sec:optimal:subpacket}
Consider a single server network with constraint $\rm{LSP_U}$ as in~\cite{Y:D:SP-LFR}. For any $(K,F,Z,S)$, the scheme proposed in~\cite{Y:D:SP-LFR} from PDA $\mathbf{A}$ achieves the memory-load pair $\big(\overline{M}_{\mathbf{A}},\overline{R}_{\mathbf{A}}\big)=\big(1+\frac{Z(N-1)}{F},\frac{S}{F}\big)$. The following conclusion was proved in~\cite{Y:D:SP-LFR}.
\begin{lemma}[{From~\cite[Theorem 2]{Y:D:SP-LFR}}]\label{lemma:single:PDA} Given a $(K,F,Z,S)$ PDA $\mathbf{A}$, 
if the associated  scheme for the single server system with constraint $\rm{LSP_U}$  achieves a memory-load pair $(\overline{M}_{\mathbf{A}},\overline{R}_{\mathbf{A}})$,
then necessarily
\begin{IEEEeqnarray}{c}
\overline{R}_{\mathbf{A}}\geq \frac{K(N-\overline{M}_{\mathbf{A}})}{N-1+K(\overline{M}_{\mathbf{A}}-1)}.\label{eqn:PDA:bound}
\end{IEEEeqnarray}
In particular, the memory-load pairs $\{(\overline{M}_{t},\overline{R}_{t}):t\in[0:K]\}$ given in~\eqref{eqn:MRt:single} satisfy~\eqref{eqn:PDA:bound} with  equality. Moreover, if $\overline{M}_{\mathbf{A}}=\overline{M}_{t}$ and $\overline{R}_{\mathbf{A}}=\overline{R}_{t}$ for some $t\in[0:K]$, then $F\geq{K\choose t}$.
\end{lemma}

Now consider a $(K,F,Z,S)$ PDA $\mathbf{A}$. Assume that the associated RSP-LFR scheme achieves the memory-load pair $(M,R)=(M_{\mathbf{A}},R_{\mathbf{A}})$, then 
\begin{subequations}
\begin{IEEEeqnarray}{rCl}
R&=&R_{\mathbf{A}}
=\frac{H}{L}\cdot \overline{R}_{\mathbf{A}}\label{eqn:R:replace}\\
&\geq&\frac{H}{L}\cdot \frac{K(N-\overline{M}_{\mathbf{A}})}{N-1+K(\overline{M}_{\mathbf{A}}-1)}\label{eqn:lemma}\\
&=& \frac{HK(N-M)}{L(N-1+K(M-1))}\label{eqn:Meq}\\
&=&\frac{H(K-x)}{L(1+x)}\bigg|_{x=K\frac{M-1}{N-1}},
\end{IEEEeqnarray}
\end{subequations}
where:
\eqref{eqn:R:replace} follows from Remark~\ref{rem:comparison}; 
\eqref{eqn:lemma} follows from~\eqref{eqn:PDA:bound}; 
and~\eqref{eqn:Meq} follows from the fact $M=M_{\mathbf{A}}=\overline{M}_{\mathbf{A}}$ by Remark~\ref{rem:comparison}.
Therefore, we proved~\eqref{eqn:PDA:bd}. 

The fact that  memory pairs $\{(M_t,R_t):t\in[0:K]\}$ satisfy~\eqref{eqn:PDA:bd}  with equality can be verified trivially. 
Moreover, if $M=M_{\mathbf{A}}=M_t$ and $R=R_{\mathbf{A}}=R_t$, then $\overline{M}_{\mathbf{A}}=\overline{M}_{t}$ and $\overline{R}_{\mathbf{A}}=\overline{R}_{t}$, by the facts $M_t=\overline{M}_{t}, R_t=\frac{H}{L}\cdot \overline{R}_{t}$ and  Remark~\ref{rem:comparison}. 
Therefore, by Lemma \ref{lemma:single:PDA}, it must hold that $F\geq {K\choose t}$. Thus, by Theorem~\ref{thm:PDA}, the subpacketization of the RSP-LFR scheme is at least $L{K\choose t}$.

\section{Improved  Load-Memory Tradeoffs in Less Constrained Systems}\label{sec:degraded}
The basic idea  for improving the load-memory tradeoff in less constrained systems is that in the case the security condition~\eqref{eqn:security} is not imposed (i.e., the constraint $\mathcal{C}$ does not contain $\rm{S}$),  some redundant signals may be removed when $N\leq K$ as in~\cite{Yu2019ExactTradeoff,Kai2020LinearFunction}. Notice that in such less constrained systems,   $R_{\mathcal{C}}^*(M)$ is defined on $M\in[0,N]$. 

Consider a fixed MAN-PDA $\mathbf{A}_t$ in~\eqref{eq:MAN:aij}, where $F={K\choose t}$ and $S={K\choose t+1}$.  Notice that each row of $\mathbf{A}_t$ is assocated to a subset of size $t$, i.e., for any given $\sd{a}\in\mathbb{F}_q^N$ and $l\in[L]$ or $h\in[H]$, each linear combination of files $W_{\sd{a},l,u}$ or $\overline{W}_{\sd{a}, h, u}$ is associated to the subset $\mathcal{T}_u\subseteq[K]$. For notational  simplicity, in this section, for each $u\in [{K\choose t}]$, denote
\begin{IEEEeqnarray}{c}
W_{\sd{a},l,\mathcal{T}_u}:=W_{\sd{a},l,u},\quad\overline{W}_{\sd{a},h,\mathcal{T}_{u}}:=\overline{W}_{\sd{a},h,u}.\label{eqn:WT}
\end{IEEEeqnarray}
Moreover, each signal $Y_{l,s}$ or $\overline{Y}_{h,s}$ is associated to a subset $\mathcal{J}\subseteq [K]$ of size $t+1$, i.e., the subset $\mathcal{J}$ such that $s=\kappa_{t+1}(\mathcal{J})$. Denote
\begin{IEEEeqnarray}{c}
Y_{l,\mathcal{J}}:=Y_{l,\kappa_{t+1}(\mathcal{J})},\quad\overline{Y}_{h,\mathcal{J}}:=\overline{Y}_{h,\kappa_{t+1}(\mathcal{J})}.\label{eqn:YJ}
\end{IEEEeqnarray}
In RP-LFR, RP-FR and R-LFR  systems,  the security condition~\eqref{eqn:security} is not imposed. Thus, the security keys can be dropped, i.e., instead of generating the random variables in~\eqref{eqn:random:V} we set
\begin{IEEEeqnarray}{c}
V_{l,s}=\mathbf{0},\quad \forall \,l\in[L],s\in[S].\label{eqn:drop:security}
\end{IEEEeqnarray}
 Therefore, with notations as in~\eqref{eqn:WT} and~\eqref{eqn:YJ}, by~\eqref{eqn:Yhs} and~\eqref{eqn:information:Yls}, we have
\begin{IEEEeqnarray}{c}
Y_{l,\mathcal{J}}=\sum_{j\in\mathcal{J}}W_{\sd{q}_j,l,\mathcal{J}\backslash\{j\}},\quad
 \overline{Y}_{h,\mathcal{J}}=\sum_{j\in\mathcal{J}}\overline{W}_{\sd{q}_j,h,\mathcal{J}\backslash\{j\}},\IEEEeqnarraynumspace
\end{IEEEeqnarray}
where $(\overline{Y}_{1,\mathcal{J}},\ldots,\overline{Y}_{H,\mathcal{J}})$ is the MDS coded version of $(Y_{1,\mathcal{J}},\ldots,Y_{L,\mathcal{J}})$ with generator matrix $G$. 

\subsection{Improved Tradeoff in RP-LFR System (Proof of Theorem~\ref{thm:RP-LFR})}\label{subsec:RP-LFR}
In RP-LFR system, the robust correctness, user-side and server-side privacy conditions are guarantted for all LFR demands. Notice that, the point $(0,\frac{HN}{L})$ can be achieved by trivially transmitting the whole coded subfiles $\overline{W}_{[N],h}$ to the users for any server $h\in[H]$. The point $\big(M_K^{\mathrm{LP}},R_K^{\mathrm{LP}})=(N,0)$ can be achieved by trivially storing all the $N$ files at each user. In the following, we describe the scheme achieving the point $\big(M_{t}^{\mathrm{LP}},R_{t}^{\mathrm{LP}})$ in~\eqref{eqn:MR:RPL} for each fixed $t\in[0:K-1]$. The lower convex envelope of those points can be achieved by memory-sharing technique.

\paragraph*{Placement Phase}
The cached contents of the users are generated as in~\eqref{eqn:Zk} according to $\mathbf{A}_t$, i.e., with notations as in~\eqref{eqn:WT}, user~$k\in[K]$ caches 
\begin{subequations}\label{eqn:Zk:RP-LFR}
\begin{IEEEeqnarray}{l}
Z_k=\{W_{n,l,\mathcal{T}}:n\in[N],l\in[L], \mathcal{T}\subseteq[K],\notag\\
\quad |\mathcal{T}|=t,k\in\mathcal{T}\}\label{eqn:Zk:1}\\
\cup\{W_{\sd{p}_k,l,\mathcal{T}}:l\in[L], \mathcal{T}\subseteq[K],|\mathcal{T}|=t,k\notin\mathcal{T}\}\IEEEeqnarraynumspace\label{eqn:Zk:2}\\
\cup\{\sd{p}_k\}.\label{eqn:Zk:3}
\end{IEEEeqnarray}
\end{subequations}
\paragraph*{Delivery Phase}

The queries $\sd{q}_{[K]}$ are generated as in~\eqref{eqn:Qkh}. Let $\mathcal{I}\subseteq[K]$ be a subset such that the vectors $\sd{q}_{\mathcal{I}}$ form a maximum linear independent vector group of the vectors $\sd{q}_{[K]}$. Each server $h\in[H]$ sends 
\begin{IEEEeqnarray}{c}\label{eqn:Xh:RPL}
X_h^{\rm{LP}}=\big(\sd{q}_{[K]},\overline{Y}_{h}(\mathcal{I})\big),
\end{IEEEeqnarray}
where
\begin{IEEEeqnarray}{c}\label{eqn:YhI}
\overline{Y}_{h}(\mathcal{I})\triangleq\{\overline{Y}_{h,\mathcal{J}}:\mathcal{J}\subseteq[K],|\mathcal{J}|=t+1,\mathcal{J}\cap\mathcal{I}\neq \emptyset\}.\IEEEeqnarraynumspace
\end{IEEEeqnarray}

\paragraph*{Robust Correctness} For any fixed $\mathcal{J}\subseteq[K]$ of size $t+1$,  $(\overline{Y}_{1,\mathcal{J}},\ldots,\overline{Y}_{H,\mathcal{J}})$ is the MDS coded version of $(Y_{1,\mathcal{J}},\ldots,Y_{L,\mathcal{J}})$ with generator matrix $G$. Thus with signals from any $L$ servers, each user can decode
\begin{IEEEeqnarray}{c}
\{Y_{l,\mathcal{J}}:l\in[L],\mathcal{J}\subseteq[K],|\mathcal{J}|=t+1,\mathcal{J}\cap\mathcal{I}\neq\emptyset\}.\IEEEeqnarraynumspace
\end{IEEEeqnarray}
Moreover, for each fixed $l\in[L]$, by the results in~\cite{Kai2020LinearFunction}, the signals $\{Y_{l,\mathcal{J}}\}_{\mathcal{J}\subseteq[K],|\mathcal{J}|=t+1}$ can be decoded from the signals $\{Y_{l,\mathcal{J}}\}_{\mathcal{J}\subseteq[K],|\mathcal{J}|=t+1,\mathcal{J}\cap\mathcal{I}\neq\emptyset}$. As a result, each user~$k\in[K]$ can decode 
\begin{IEEEeqnarray}{rl}
&\{Y_{l,\mathcal{J}}\,:\,l\in[L],\mathcal{J}\subseteq[K],|\mathcal{J}|=t+1\}\notag\\
&=\{Y_{l,s}:l\in[L],s\in[S]\},
\end{IEEEeqnarray}
i.e., all the signals in~\eqref{eqn:information:Yls}. By continue with the same arguments following~\eqref{eqn:information:Yls}, each user can correctly decode its demanded linear combination of the files. 

\paragraph*{User/Server-side Privacy}
The proof that the scheme guarantees the server-side and user-side privacy conditions follow the same line of reasoning as in~\eqref{eqn:user:privacy:proof}  and \eqref{eqn:server:privacy:proof}, respectively.  

\paragraph*{Performance} By~\eqref{eqn:Zk:RP-LFR}, each user stores $NL{K-1\choose t-1}+L{K-1\choose t}$ packets, each of size $\frac{B}{L{K\choose t}}$, and a vector $\sd{p}_k\in\mathbb{F}_q^N$ of
 length $N$. The needed memory size is given by 
\begin{subequations}
\begin{IEEEeqnarray}{rl}
M_t^{\mathrm{LP}}=&\inf_{B\in\mathbb{N}^+}\frac{1}{B}\Big(\frac{B\big(NL{K-1\choose t-1}+L{K-1\choose t}\big)}{L{K\choose t}}+N\Big)\IEEEeqnarraynumspace\\
=&1+\frac{t(N-1)}{K}.
\end{IEEEeqnarray}
\end{subequations}
Let $\mathrm{rank}_q(\sd{q}_{[K]})$ be the rank of vectors $\sd{q}_{[K]}$, i.e., the cardinality of $\mathcal{I}$.  By~\eqref{eqn:Xh:RPL} and~\eqref{eqn:YhI}, each server sends ${K\choose t+1}-{K-\mathrm{rank}_q(\sd{q}_{[K]})\choose t+1}$ packets, and $K$ vectors of length $N$. Notice that  the worst case is $\mathrm{rank}_q(\sd{q}_{[K]})=\min\{N,K\}$, therefore, the load is given by 
\begin{subequations}
\begin{IEEEeqnarray}{rl}
R_t^{\mathrm{LP}}
=&\inf_{B\in\mathbb{N}^+}\frac{1}{B}\Big(\frac{HB\big({K\choose t+1}-{K-\min\{K,N\}\choose t+1}\big)}{L{K\choose t}}+NK\Big)\IEEEeqnarraynumspace\\
=&\frac{H\big({K\choose t+1}-{K-\min\{K,N\}\choose t+1}\big)}{L{K\choose t}}.
\end{IEEEeqnarray} 
\end{subequations}
\paragraph*{Gap Result} Let $\overline{R}_{\mathrm{LP_U}}(M)$ be the load-memory tradeoff achieved by the scheme in~\cite{Y:D:Privacy} in the single server case, where user-side privacy is guaranteed for all LFR demands, which is given by the lower convex envelope of the point $(0,N)$ and the following points
\begin{IEEEeqnarray}{rl}
&\big(\overline{M}_{t}^{\mathrm{LP_U}},\overline{R}_{t}^{\mathrm{LP_U}})=\notag\\
&\bigg(1+\frac{t(N-1)}{K},\frac{{K\choose t+1}-{K-\min\{K,N\}\choose t+1}}{{K\choose t}}\bigg),
\end{IEEEeqnarray} 
where $t\in[0:K]$.
Notice that, for the corner points with $M=0$ and $M\in \{M_t^{\mathrm{LP_U}}:t\in[0:K]\}$, it always hold 
\begin{IEEEeqnarray}{c}
R_{\mathrm{LP}}(M)=\frac{H}{L}\cdot \overline{R}_{\mathrm{LP_U}}(M).\label{eqn:R-M:RPL}
\end{IEEEeqnarray}
 Since the corner points coincide on $M$,~\eqref{eqn:R-M:RPL} hold for all $M\in[0,N]$.  Moreover,
\begin{IEEEeqnarray}{c}
\frac{R_{\rm{LP}}(M)}{R_{\rm{LP}}^*(M)}\overset{(a)}{\leq} \frac{\overline{R}_{\rm{LP_U}}(M)}{\overline{R}_{\rm{LP}}^*(M)}\overset{(b)}{\leq}  \frac{\overline{R}_{\rm{LP_U}}(M)}{\overline{R}_{\rm{LP_U}}^*(M)},\label{bound:chain}
\end{IEEEeqnarray}
where:
 $(a)$ follows from Lemma \ref{lemma:bound} and~\eqref{eqn:R-M:RPL}; and (b) follows from the fact $\overline{R}_{\rm{LP}}^*(M)\geq \overline{R}_{\rm{LP_U}}^*(M)$, since the constraint $\rm{LP}$ is stronger than the constraint $\rm{LP_U}$.
Then the gap result directly follows from bound for $\frac{\overline{R}_{\rm{LP_U}}(M)}{\overline{R}_{\rm{LP_U}}^*(M)}$ in \cite[Theorem 6]{Y:D:Privacy}.

\subsection{Improvement in RP-FR System (Proof of Theorem~\ref{thm:RP-FR})}
In RP-FR system, the robust correctness, user-side and server-side privacy conditions are guaranteed for all FR demands. The proof of Theorem~\ref{thm:RP-FR} follows similarly to the proof of Theorem~\ref{thm:RP-LFR} in Section~\ref{subsec:RP-LFR}, with the following distinctions.

\paragraph*{Placement Phase} Instead of generating $\sd{p}_1,\ldots,\sd{p}_K$ uniformly from $\mathbb{F}_q^N$, we let $\sd{p}_1,\ldots,\sd{p}_K$ generated uniformly from 
$
\big\{(x_1,\ldots,x_N)^{\top}\in\mathbb{F}_q^N:\sum_{n\in[N]}x_n=q-1\big\}
$.
\paragraph*{Performance} Since the queries $\sd{q}_1,\ldots,\sd{q}_K$ are generated as in~\eqref{eqn:Qkh} and the demands $\sd{d}_1,\ldots,\sd{d}_K\in\{\sd{e}_1,\ldots,\sd{e}_N\}$, the queries are uniformly distributed over the $N-1$ dimensional subspace 
$
\big\{(x_1,\ldots,x_N)^{\top}\in\mathbb{F}_q^N:\sum_{n\in[N]}x_n=0\big\}
$.
Thus, in the worst case, $\mathrm{rank}_q(\sd{q}_{[K]})=\min\{K,N-1\}$. As a result, the achieved memory-load pair $(M_t^{\mathrm{FP}}, R_t^{\mathrm{FP}})$ is given by 
\begin{IEEEeqnarray}{rl}
&(M_t^{\mathrm{FP}}, R_t^{\mathrm{FP}})=\notag\\
 &\bigg(1+\frac{t(N-1)}{K},\frac{H\big({K\choose t+1}-{K-\min\{K,N-1\}\choose t+1}\big)}{L{K\choose t}}\bigg).\IEEEeqnarraynumspace\label{eqn:MR:RPF}
\end{IEEEeqnarray}
\paragraph*{Gap Result}  Let $\overline{R}_{\rm{FP_U}}(M)$ be the lower convex envelope of the point $(0,N)$ and  points $\big\{\big(\overline{M}_t^{\rm{FP_U}},\overline{R}_t^{\rm{FP_U}}\big):t\in[0:K]\big\}$,
 where
\begin{IEEEeqnarray}{rl}
&(\overline{M}_{t}^{\mathrm{FP_U}},\overline{R}_{t}^{\mathrm{FP_U}})=\notag\\
 &\bigg(1+\frac{t(N-1)}{K},\frac{{K\choose t+1}-{K-\min\{K,N-1\}\choose t+1}}{{K\choose t}}\bigg),\IEEEeqnarraynumspace
\end{IEEEeqnarray} 
which is proved to be achievable in the single server case for all FR demands in \cite[Theorem 1]{Y:D:Privacy}. 
Following the same line of reasoning as to obtain~\eqref{bound:chain}, we have
$
R_{\rm{FP}}(M)=\frac{H}{L}\cdot\overline{R}_{\rm{FP_U}}(M)
$ for all $M\in[0,N]$, and 
\begin{IEEEeqnarray}{c}
\frac{R_{\rm{FP}}(M)}{R_{\rm FP}^*(M)}\leq \frac{\overline{R}_{\rm{FP_U}}(M)}{\overline{R}_{\rm{FP}}^*(M)}\leq  \frac{\overline{R}_{\rm{FP_U}}(M)}{\overline{R}_{\rm{FP_U}}^*(M)}.
\end{IEEEeqnarray}
Then gap result directly follows from the upper bound for $\frac{\overline{R}_{\rm{LP_U}}(M)}{\overline{R}_{\mathrm{LP_U}}^*(M)}$ in \cite[Theorem 5]{Y:D:Privacy}.

\subsection{Improvement in R-LFR System (Proof of Theorem~\ref{thm:R-LFR})}
In the R-LFR system, only the robust correctness condition must be guaranteed for all LFR demands. As a result,  in addition to dropping the security keys (see~\eqref{eqn:drop:security}), the privacy keys can also be dropped, i.e., set to zero.  In particular, the stored contents in~\eqref{eqn:Zk:2} and~\eqref{eqn:Zk:3} can be dropped, i.e., set to zero. The correctness can be easy verified by setting $\sd{p}_1=\ldots=\sd{p}_K=\mathbf{0}$ and following the same line of reasoning as in Section~\ref{subsec:RP-LFR}. The distinctions are in performance and gap results.
\paragraph*{Performance} In the modified scheme for R-LFR system, only the contents in~\eqref{eqn:Zk:1} are stored. The delivered signals are the same as in~\eqref{eqn:Xh:RPL}. Thus, the achieved memory-load pair is given by 
\begin{IEEEeqnarray}{c}\label{eqn:MR:RL}
(M_t^{\mathrm{L}},R_t^{\mathrm{L}})=\bigg(\frac{tN}{K},\frac{H\big({K\choose t+1}-{K-\min\{K,N\}\choose t+1}\big)}{L{K\choose t}}\bigg),\IEEEeqnarraynumspace
\end{IEEEeqnarray}
where $t\in[0:K]$. The lower convex envelope of those points can be achieved by memory-sharing. 

\paragraph*{Gap Result}  Let $\overline{R}_{\rm F}(M)$ be the lower convex envelope of the points $\big\{\big(\overline{M}_t^{\rm F},\overline{R}_t^{\rm F}\big):t\in[0:K]\big\}$ where
\begin{IEEEeqnarray}{c}
(\overline{M}_{t}^{\mathrm{F}},\overline{R}_{t}^{\mathrm{F}})=\bigg(\frac{tN}{K}, \frac{{K\choose t+1}-{K-\min\{K,N\}\choose t+1}}{{K\choose t}}\bigg),\IEEEeqnarraynumspace\label{eqn:MR:SRL}
\end{IEEEeqnarray} 
which is proved to be achievable in the single server case for all FR demands in~\cite{Yu2019ExactTradeoff}. 
Following the same line of reasoning as to obtain~\eqref{bound:chain}, we have
$
R_{\rm{L}}(M)=\frac{H}{L}\cdot\overline{R}_{\rm{F}}(M)
$ for all $M\in[0,N]$, and 
\begin{IEEEeqnarray}{c}
\frac{R_{\rm{L}}(M)}{R_{\rm L}^*(M)}\leq \frac{\overline{R}_{\rm{F}}(M)}{\overline{R}_{\rm{L}}^*(M)}\leq  \frac{\overline{R}_{\rm{F}}(M)}{\overline{R}_{\rm{F}}^*(M)}.
\end{IEEEeqnarray}
Then the gap result follows from the upper bound for $ \frac{\overline{R}_{\rm F}(M)}{\overline{R}_{\mathrm{F}}^*(M)}$ in \cite[Theorem 1]{YuFactor2}.

\section{Numerical Results}\label{sec:numerical}
 In Fig.~\ref{fig:2}, we plot the achievable memory-load tradeoff  under different constrains (Theorems \ref{thm:MAN}, \ref{thm:RP-LFR}, \ref{thm:RP-FR} and \ref{thm:R-LFR})  for the three regimes:
\begin{enumerate}
\item [a)] $N\geq \frac{K+1+\sqrt{3K^2+1}}{2}$, see Fig.~\ref{fig2:a};
\item [b)]$K<N<\frac{K+1+\sqrt{3K^2+1}}{2}$,  see Fig.~\ref{fig2:b}; and
\item[c)] $N\leq K$, see Fig.~\ref{fig2:c}.
\end{enumerate}
We choose parameters  $(N,K,L,H)=(30,10,15,20),(25,20,15,20), (10,30,15,20)$, respectively. 
From the figures, we observe:
\begin{figure}[htbp] \centering
\subfigure[$(N,K,L,H)=(30,10,15,20)$] {
 \label{fig2:a}
\includegraphics[width=0.85\columnwidth]{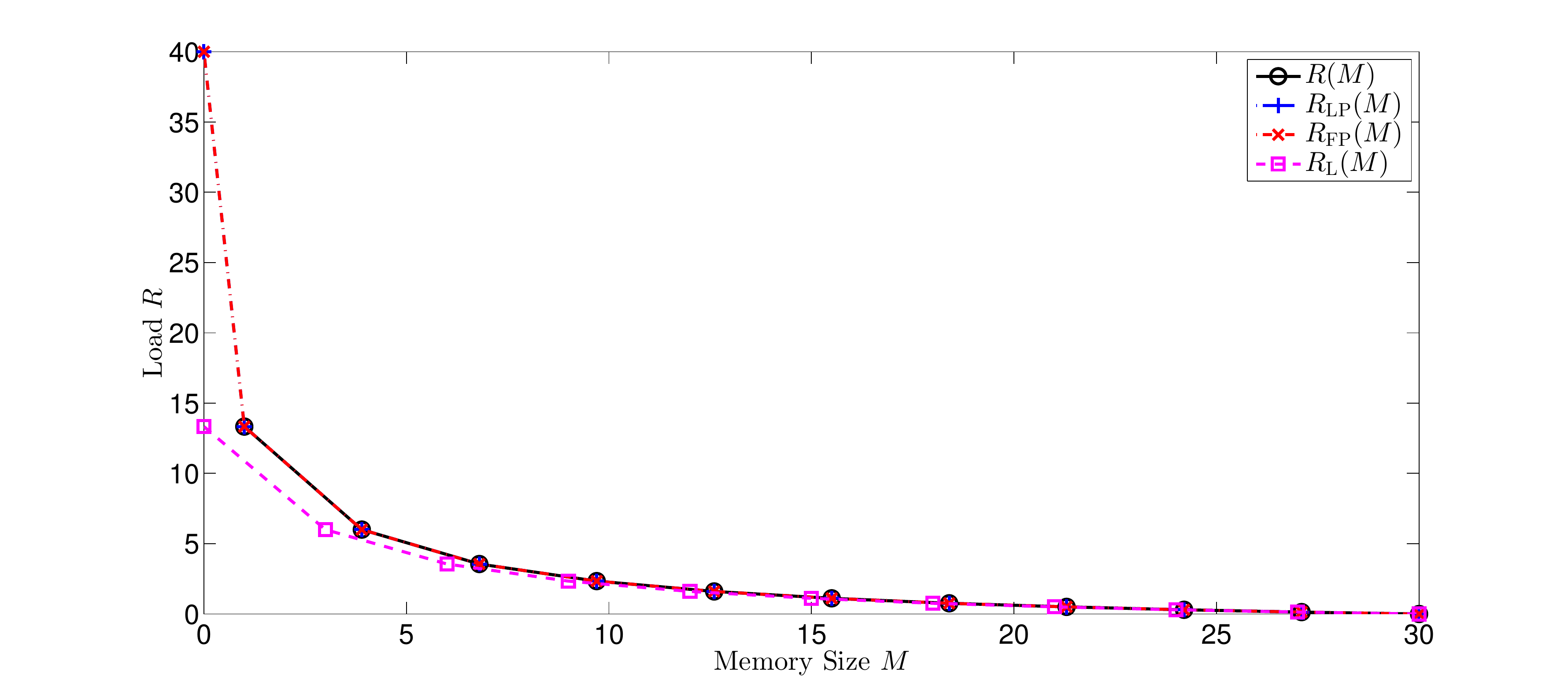}
}
\subfigure[$(N,K,L,H)=(25,20,15,20)$] {
 \label{fig2:b}
\includegraphics[width=0.85\columnwidth]{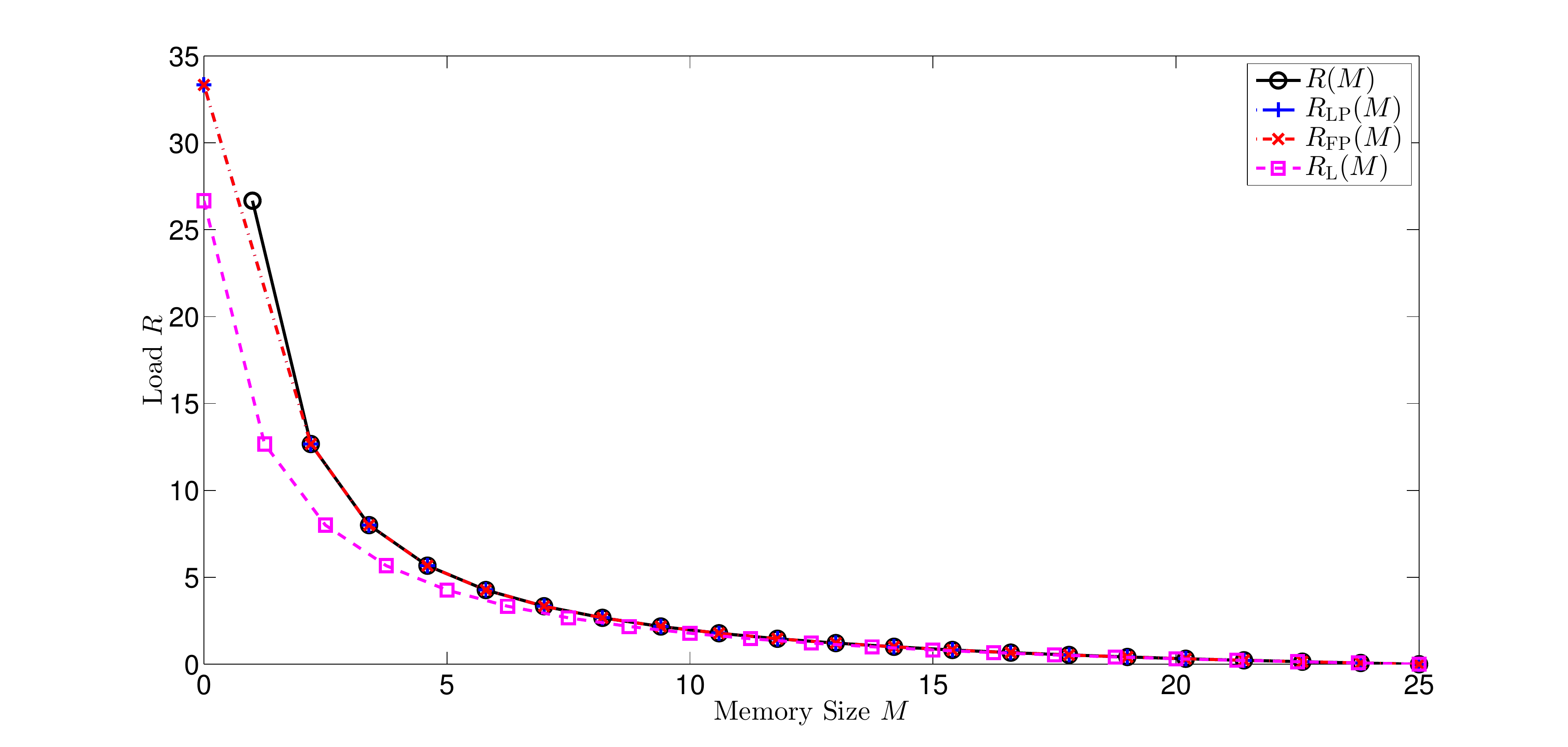}}
\subfigure[$(N,K,L,H)=(10,30,15,20)$] {
\label{fig2:c}
\includegraphics[width=0.85\columnwidth]{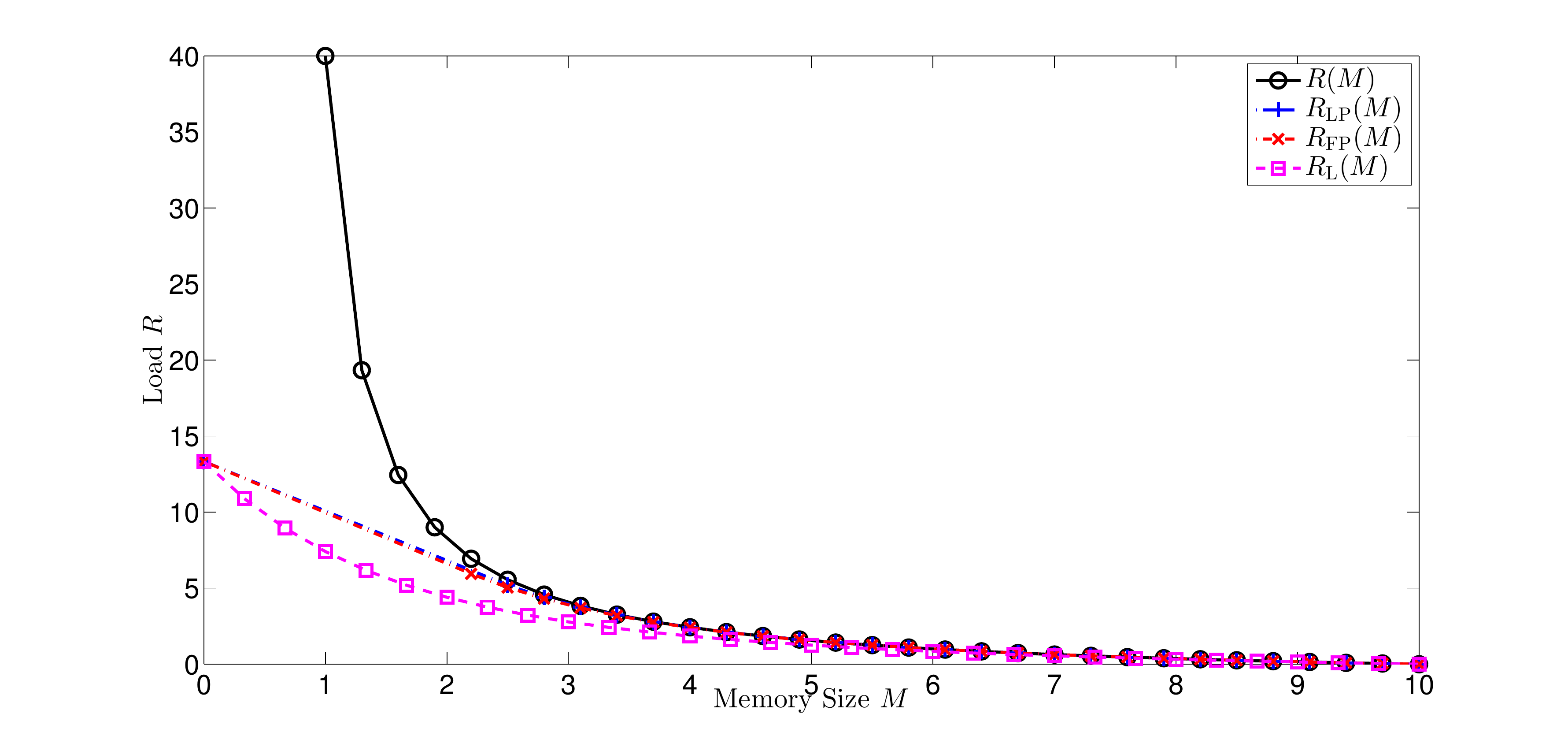}
}
\caption{Load-memory tradeoffs for robust systems  (a) $N\geq \frac{K+1+\sqrt{3K^2+1}}{2}$; (b) $ K<N<\frac{K+1+\sqrt{3K^2+1}}{2}$; (c) $N\leq K$. }
\label{fig:2}
\end{figure}

1) For $N\geq \frac{K+1+\sqrt{3K^2+1}}{2}$ (Fig. \ref{fig2:a}), the MAN-PDA based scheme in the  RSP-LFR system achieves the same tradeoff as that in the RP-LFR and RP-FR systems on the interval $M\in[1,N]$. This is because: 
\begin{enumerate}
\item [i)] there is no redundant signals to be removed in RP-LFR or RP-FR;
\item [ii)] the privacy keys and security keys are stored in the superposition form;
\item [iii)]  the lower convex envelope of $(0,N)$ and $\{(M_t,R_t):t\in[0:K]\}$ are formed by connecting $(0,\frac{HN}{L})$ and $(M_0,R_0),(M_1,R_1),\ldots,(M_K,R_K)$ sequentially. This can be verified by letting the slope of the line connecting $(0,\frac{HN}{L})$ and $(M_0,R_0)$ be no larger than the slope of connecting $(M_0,R_0)$ and $(M_1,R_1)$, i.e., 
\begin{IEEEeqnarray}{c}
\frac{R_0-HN/L}{M_0-0}\leq \frac{R_1-R_0}{M_1-M_0},\label{eqn:slope}
\end{IEEEeqnarray}
which indicates that $N$ should satisfy $N\geq \frac{K+1+\sqrt{3K^2+1}}{2}$. 
\end{enumerate}
The improved tradeoff in R-LFR system is due to the saved memory for keys for the regime $M\in[1,N]$, and there is no need to guarantee privacy by  sending all coded files at $M=0$ (i.e., the point $(0,K)$ is achievable in R-LFR system). 

2) For $K<N< \frac{K+1+\sqrt{3K^2+1}}{2}$ (Fig. \ref{fig2:b}),  similar phenomena are observed as in the case $N\geq \frac{K+1+\sqrt{3K^2+1}}{2}$, except that now there is slightly improvement in RP-LFR and RP-FR systems over the RSP-LFR system in the interval $M\in\big[1,1+\frac{N-1}{K}\big]$. This improvement comes from taking the lower convex envelope with the additional point $(0,\frac{HN}{L})$ (observe that~\eqref{eqn:slope} does not hold). Notice that for the case $N>K$ (Fig. \ref{fig2:a} and \ref{fig2:b}), all the tradeoffs are proved to be within a constant multiplicative gap of the optimal tradeoff in their respective setups.

3) For the case $N\leq K$ (Fig.~\ref{fig2:c}), the tradeoff in RP-LFR and RP-FR systems significantly smaller than that in the LSP-LFR system for small $M$ regime, because:
      \begin{enumerate}
        \item[i)] The trivial point $(M,R)=(0,\frac{HN}{L})$ can be achieved, and thus memory-sharing the other points with this point increases the performance.
        \item[ii)] For $M\in\{M_t: t\in[0:K-N]\}$, some redundant signals are removed in RP-LFR and RP-FR, similarly to~\cite{Kai2020LinearFunction,Yu2019ExactTradeoff}.
      \end{enumerate}
In this case, due to the use of security keys in the RSP-LFR system, the counterpart of redundant signals in RP-LFR and RP-FR system can not be obtained from the counterpart of the transmitted signals. Notice that, the tradeoff in RP-FR is slightly better than that in the RP-LFR system, since the number of removed redundant signals in RP-FR system is ${K-N+1\choose t+1}$, which is larger than that in the RP-LFR system ${K-N\choose t+1}$. The improvement in the R-LFR system over RP-LFR/RP-FR systems comes from the saved memory size for privacy keys.

\section{Conclusion}\label{sec:conclusion}
A PDA-based key superposition RSP-LFR scheme is proposed 
for MDS distributed storage systems that simultaneously
guarantees content security against a wiretapper having access
to the delivery signals and demand privacy against both servers
and colluding users. The load-memory tradeoff turns out to be 
the single-server one scaled by
the inverse of the rate of the MDS code in order to guarantee 
robustness against link/server failures. 
The performance of MAN-PDA-based RSP-LFR scheme is showed to be to within a
multiplicative gap of at most eight from optimal in all regimes, 
except for small memory regime with less files than users. 
Moreover, in three less restrictive systems without the security constraint (i.e., RP-LFR, RP-FR, and R-LFR systems),  some redundant signals can be removed to further improve the load-memory tradeoff, which are proved to be within a constant multiplicative gap of the optimal  tradeoff in their respective setups.

\end{document}